\documentclass[twocolumn,pre,showpacs]{revtex4}
\usepackage{graphicx}
\usepackage{amsmath}
\usepackage{amsfonts}
\usepackage{amssymb}
\usepackage{color}
\usepackage{float}
\usepackage{mathrsfs}

\newcommand {\ra}{\rightarrow}

\newcommand {\pt}{\partial}

\begin{document}
\title{Strongly interacting soliton gas and formation of rogue waves}
\author{A.\,A.~Gelash$^{1,2}$}\email{gelash@srd.nsu.ru}
\author{D.\,S.~Agafontsev$^{1,3}$}
\affiliation{$^{1}$Novosibirsk State University, Novosibirsk, 630090, Russia}
\affiliation{$^{2}$Institute of Thermophysics, SB RAS, Novosibirsk, 630090, Russia}
\affiliation{$^{3}$P.\,P. Shirshov Institute of Oceanology, RAS, 117218, Moscow, Russia}
\pacs{05.45.Yv,02.30.Ik,42.81.Dp}

\begin{abstract}
We study numerically the properties of (statistically) homogeneous soliton gas depending on soliton density (proportional to number of solitons per unit length) and soliton velocities, in the framework of the focusing one-dimensional Nonlinear Schr{\"o}dinger (NLS) equation. 
In order to model such gas we use $N$-soliton solutions ($N$-SS) with $N\sim 100$, which we generate with specific implementation of the dressing method combined with $100$-digits arithmetics. 
We examine the major statistical characteristics, in particular the kinetic and potential energies, the kurtosis, the wave-action spectrum and the probability density function (PDF) of wave intensity. 

We show that in the case of small soliton density the kinetic and potential energies, as well as the kurtosis, are very well described by the analytical relations derived without taking into account soliton interactions. 
With increasing soliton density and velocities, soliton interactions enhance, and we observe increasing deviations from these relations leading to increased absolute values for all of these three characteristics. 
The wave-action spectrum is smooth, decays close to exponentially at large wavenumbers and widens with increasing soliton density and velocities. 
The PDF of wave intensity deviates from the exponential (Rayleigh) PDF drastically for rarefied soliton gas, transforming much closer to it at densities corresponding to essential interaction between the solitons. 
Rogue waves emerging in soliton gas are multi-soliton collisions, and yet some of them have spatial profiles very similar to those of the Peregrine solutions of different orders. 
We present example of three-soliton collision, for which even the temporal behavior of the maximal amplitude is very well approximated by the Peregrine solution of the second order. 
\end{abstract}

\maketitle


\section{Introduction}
\label{Sec:Intro}

Statistical behavior of nonlinear integrable systems, called in general \textit{integrable turbulence}~\cite{zakharov2009turbulence}, is a rapidly developing area of theoretical and experimental studies, as illustrated by recent publications~\cite{randoux2014intermittency,walczak2015optical,agafontsev2015integrable,agafontsev2016integrable}. 
From one hand, up to a certain degree of accuracy many physical systems can be described with nonlinear integrable mathematical models. 
In comparison with nonintegrable models, the corresponding integrable equations demonstrate significantly different statistical properties, see e.g.~\cite{suret2011wave,picozzi2014optical}. 
On the other hand, an integrable system allows transformation to the so-called \textit{scattering data}, which is in one-to-one correspondence with the wavefield and, similarly to the Fourier harmonics in the linear wave theory, change trivially during the motion. 
With numerical methods, see e.g.~\cite{randoux2016inverse,akhmediev2016breather}, the scattering data can be partly analyzed, what may bring some insights into the dynamical behavior. 
Another distinctive feature of an integrable system is the conservation of infinite series of invariants, so that different types of initial conditions are characterized by different sets of integrals of motion and, during the evolution, demonstrate different statistical properties, see e.g.~\cite{walczak2015optical,agafontsev2015integrable,agafontsev2016integrable}. 

In the present paper we examine integrable turbulence using controlled initial conditions, in the sense that we construct these initial conditions from known scattering data. 
In contrast to other studies, this gives us exact knowledge which nonlinear objects interact during the evolution, for instance, when a rogue wave appear. 
As a model, we consider one-dimensional Nonlinear Schr{\"o}dinger (NLS) equation of the focusing type with initial conditions in the form of $N$-soliton solutions ($N$-SS), with $N$ of order $100$. 
Our methods allow generation of sufficiently dense $N$-SS with essential interaction between the solitons, in contrast to rarefied multi-soliton solutions analyzed, e.g., in~\cite{pelinovsky2013two,dutykh2014numerical,shurgalina2016nonlinear} for KdV and mKdV equations.
We believe that our approach can also be used to examine turbulence governed by other integrable equations and developing from other types of initial conditions, e.g. containing nonlinear dispersive waves and different types of breathers, see~\cite{frumin2017new,gelash2018formation}. 

For spatially localized wavefield, the scattering data consists of discrete (solitons) and continuous (nonlinear dispersive waves) parts of eigenvalue spectrum, which is calculated for specific auxiliary linear system. 
At the first step in our study, we generate an ensemble of multiple realizations of scattering data, with each realization containing $N$ discrete eigenvalues and $N$ complex coefficients. 
Such scattering data corresponds to $N$-SS. 
Then, we find the wavefield for $N$-SS from this data, what for $N\sim 100$ is made possible by specific implementation of the dressing method applied numerically with $100$-digits precision. 
To our knowledge, multi-soliton solutions containing so many solitons were not generated by anyone else before. 
The generating procedure is very expensive from the computational point of view and returns a wavefield where solitons are distributed unevenly over the spatial dimension. 
This is why we then put multi-soliton solutions in a periodic box $L$ and examine their evolution with the direct numerical simulation. 
After some time, the solitons distribute over the box uniformly and the system arrives to the statistically steady state where its statistical characteristics do not change anymore. 
We use this state as a model of statistically homogeneous soliton gas of density $\propto N/L$ in an infinite space; we confirmed that for large enough number of solitons $N$ and period $L$ our results depend on them only in combination $N/L$. 
We study the major statistical characteristics of such soliton gas, namely the kinetic and potential energies, the kurtosis, the wave-action spectrum, and the probability density function (PDF) of wave intensity, averaging the results over the ensemble of initial conditions. 
The described methods allow us to explore statistical behavior of sufficiently dense soliton gas with randomized soliton amplitudes and velocities. 

Formally, when we simulate evolution of multi-soliton solutions in a periodic box, we change a localized wavefield generated from discrete scattering data by a periodic one, which corresponds to a finite-gap scattering data~\cite{belokolos1994algebro}. 
However, since the widths of the solitons are by two-three orders of magnitude smaller than the size of the box, these eigenvalue gaps are very narrow~\cite{belokolos1994algebro,bobenko2011computational} and we neglect this difference. 
When generating $N$-SS from the scattering data, we also limit spatial positions of solitons in such a way to ensure that the intensity of the wavefield near the edges of the periodic box is of round-off order. 
This allows us to neglect edge effects as well. 
We confirmed that after a long evolution the first $10$ integrals of motion are conserved with a very good accuracy and the resulting eigenvalue spectrum virtually coincides with the initial discrete eigenvalues. 

In the present study we focus on the statistical description of large waves -- the so-called rogue waves. 
For this purpose we examine the PDF $\mathcal{P}(I)$ of relative wave intensity $I=|\psi|^{2}/\langle\overline{|\psi|^{2}}\rangle$. 
Here $\langle\overline{|\psi|^{2}}\rangle$ is ensemble and space average of wavefield intensity $|\psi|^{2}$, so that small waves correspond to $I\ll 1$, the moderate ones to $I\sim 1$, and the large ones to $I\gg 1$, with the formal definition of rogue waves $I>8$, see e.g.~\cite{kharif2003physical, dysthe2008oceanic}. 
The PDF is normalized as 
$$
\int_{0}^{+\infty} \mathcal{P}(I)\,dI = 1.
$$
If the wavefield is a superposition of a multitude of uncorrelated linear waves with random phases and amplitudes satisfying the central limit theorem, then its PDF is the exponential distribution (or Rayleigh one after certain change of variables, see e.g.~\cite{nazarenko2011wave}), 
\begin{equation}\label{Rayleigh}
\mathcal{P}_{R}(I) = e^{-I}. 
\end{equation}
When evolution is governed by linear equations, this superposition stays uncorrelated and its PDF remains exponential. 
Nonlinear evolution may introduce correlation, which in turn may lead to enhanced appearance of large waves. 
Throughout the paper we use the exponential distribution~(\ref{Rayleigh}) as a benchmark, comparing the observed PDFs with it to make clear whether large waves appear more or less frequently in the examined system than in a linear one. 
Unless specified otherwise, we use the term PDF only in relation to relative wave intensity. 

With the described above methods, we study statistical properties of homogeneous soliton gas depending on soliton density and velocities. 
We show that, when the density is small, the kinetic and potential energies, as well as the kurtosis, are very well described by the analytical relations derived without taking into account soliton interactions. 
With increasing soliton density and velocities, soliton interactions enhance, what results in increasing deviations from these relations leading to increased absolute values for all these three characteristics. 
The wave-action spectrum for soliton gas is smooth (i.e., contrary to the condensate and cnoidal wave initial conditions~\cite{agafontsev2015integrable,agafontsev2016integrable} it doesn't contain diverging peaks), decays close to exponentially at large wavenumbers and widens with increasing soliton density and velocities. 
Compared to the cnoidal wave initial conditions, the PDF of relative wave intensity for soliton gas deviates from the exponential PDF~(\ref{Rayleigh}) much more pronouncedly and at large intensities exceeds it by orders of magnitude. 
This excess is larger for soliton gas with larger velocities. 
On the other hand, with increasing soliton density the PDF transforms closer to the exponential PDF. 
Rogue waves emerging in soliton gas are collisions of solitons, and yet some of them have spatial profiles very similar to those of the (scaled) Peregrine solutions of different orders. 
We present example of three-soliton collision, when even the temporal behavior of the maximal amplitude is very well approximated by the Peregrine solution of the second order. 
In our opinion, this fact highlights that similarity for the spatial and/or temporal behavior cannot be used to draw conclusions on rogue waves' composition and origin. 

The paper is organized as follows. 
In the next Section we give a brief introduction to the inverse scattering transform (IST) and describe how we construct multi-soliton solutions. 
In Section~\ref{Sec:NumMethods} we summarize our methods for numerical simulation of statistically homogeneous soliton gas. 
In Section~\ref{Sec:Results} we study the major characteristics of this gas, including the kinetic and potential energies, the kurtosis, the wave-action spectrum and the PDF of relative wave intensity, and examine dynamics of rogue waves. 
We finish with the Conclusions. 


\section{Inverse scattering transform and multi-soliton solutions}
\label{Sec:SolitonSolutions}

The formulation of the IST problem~\cite{zakharov1972exact}, see also~\cite{novikov1984theory,faddeev2007hamiltonian}, for the one-dimensional NLS equation of the focusing type 
\begin{equation}\label{NLSE}
i\psi_t + \frac12 \psi_{xx} + |\psi|^2 \psi=0,
\end{equation}
starts from introduction of the auxiliary Zakharov-Shabat (ZS) linear system for $2\times 2$ matrix wave function $\mathbf{\Phi}$ and complex (spectral) parameter $\lambda$, 
\begin{subequations}\label{ZH system}
\begin{eqnarray}
\label{ZH system_a}
\mathbf{\Phi}_x = \begin{pmatrix}\ -i\lambda & \psi \\ -\psi^* & i\lambda \end{pmatrix}\mathbf{\Phi},
\\
\label{ZH system_b}
\mathbf{\Phi}_t =\begin{pmatrix}\ -i\lambda^2 + \frac{i}{2} |\psi|^2 & \lambda \psi + \frac{i}{2} \psi_x \\ -\lambda \psi^* + \frac{i}{2} \psi^*_x & i\lambda^2 - \frac{i}{2} |\psi |^2 \end{pmatrix} \mathbf{\Phi},
\end{eqnarray}
\end{subequations}
from which the NLS equation~(\ref{NLSE}) is obtained as compatibility condition
$$
\mathbf{\Phi}_{xt} = \mathbf{\Phi}_{tx}.
$$
Similarly to quantum mechanics, see e.g.~\cite{landau1958quantum}, a scattering problem for $\mathbf{\Phi}$ is considered, where the wavefield $\psi(x,t)$ of the NLS equation plays the role of $x$-dependent potential. 
The so-called scattering data obtained from solution of this (direct) problem is in one-to-one correspondence with the $x$-dependency of the potential $\psi$ and can be used to reconstruct it with the inverse scattering transform. 
The key result of the IST approach is that the scattering data depends on time trivially, so that the Cauchy initial-value problem for the NLS equation can be solved formally by identifying the scattering data from the direct scattering problem, finding its evolution in time and applying the inverse scattering transform. 

The first equation of the ZS system can be rewritten in the form of eigenvalue problem, 
\begin{equation}\label{ZH system1}
\widehat{\mathcal{L}}\mathbf{\Phi} = \lambda \mathbf{\Phi},
\quad
\widehat{\mathcal{L}} = i \begin{pmatrix}\ 1 & 0 \\ 0 & -1 \end{pmatrix}\frac{\pt}{\pt x} - i\begin{pmatrix}\ 0 & \psi \\ \psi^* & 0 \end{pmatrix}.
\end{equation}
For spatially localized potentials $\psi$, eigenvalues $\lambda$ are presented by a finite number of discrete points with $\mathrm{Im}\,\lambda\neq 0$ (discrete spectrum) and the real line $\lambda\in\mathbb{R}$ (continuous spectrum). 
The scattering data consists of discrete eigenvalues $\lambda_{n}$, $n=1,...,N$, complex coefficients $C_{n}$ for each $\lambda_{n}$ and the so-called reflection coefficient $r(\xi)$,
\begin{equation}\label{ScattData}
\bigl\{ r(\xi); \quad  \lambda_n,  \quad C_n \bigr\},
\end{equation}
where $\xi$ means $\lambda$ on the real axis. 
Its time evolution is trivial, 
\begin{eqnarray}\label{ScattData(t)}
&& r(\xi,t) = r(\xi,0)e^{-2i \xi^2 t},\nonumber\\
&& \forall n: \lambda_{n}=\mathrm{const},\quad C_n (t) = C_n (0) e^{-2i \lambda_n^2 t},
\end{eqnarray}
and, in principal, allows one to find the wavefield $\psi$ for any given moment of time with the IST by solving the (integral) Gelfand--Levitan--Marchenko (GLM) equations. 
In the general case, the latter procedure can only be done numerically or asymptotically at $t\ra\infty$. 
In the present paper we consider the so-called reflectionless $r(\xi)=0$ potentials $\psi$, which represent $N$-soliton solutions of the NLS equation. 
Then, factorization of the GLM equations leads to a system of linear algebraic equations and the $N$-SS can be found in explicit form.

The simplest multi-soliton solution of the NLS equation~(\ref{NLSE}) represents $1$-SS,  
\begin{equation}\label{1-SS}
\psi_{(1)}(x,t) = a \frac{e^{i v (x-x_0) + 0.5i (a^2-v^2) t + i \theta}}{\cosh(a (x-x_0) - a v t)},
\end{equation}
and depends on four real parameters $a>0$, $v$, $x_{0}$ and $\theta$. 
The first two of them are soliton amplitude $a$ and group velocity $v$, while $x_{0}$ and $\theta$ correspond to position in space and complex phase. 
At $t=0$, the scattering data for $1$-SS is a combination of discrete eigenvalue and complex coefficient 
\begin{equation}\label{1-SS-scattering-data}
\lambda = -v/2 + i a/2, \quad C = e^{i(\theta+2\lambda x_{0})}.
\end{equation}

In order to construct $N$-SS at the initial time $t=0$, we generate $N$ pairs of discrete eigenvalues $\lambda_{n}=-v_{n}/2+i a_{n}/2$ and complex coefficients $C_{n}=e^{i(\theta_{n}+2\lambda x_{0n})}$ by using certain statistical distributions for $a_{n}$, $v_{n}$, $x_{0n}$, and $\theta_{n}$, which we discuss later in the paper. 
Here $a_{n}$, $v_{n}$, $x_{0n}$ and $\theta_{n}$ describe, respectively, amplitude, group velocity and ``approximate'' position and complex phase of the $n$-th soliton. 
Note that due to soliton interactions within the $N$-SS, the real position and complex phase of a soliton may differ considerably from $x_{0n}$ and $\theta_{n}$.
Without loss of generality, we consider $\lambda_{n}$ on the upper half-plane only, $\mathrm{Im}\,\lambda_{n}>0$, since complex-conjugated eigenvalues relate to the same class of soliton solutions. 

Then, we find the potential $\psi(x,0)$ corresponding to the generated scattering data by using the Zakharov-Mikhailov variant~\cite{zakharov1978relativistically} of the dressing method, see also~\cite{gelash2014superregular}, which can be briefly described as follows. 
Let us suppose that $\psi_{(n-1)}(x,0)$ is the $(n-1)$-SS constructed from the first $n-1$ eigenvalues $\lambda_{m}$ and complex coefficients $C_{m}$, and $\mathbf{\Phi}_{(n-1)}(x,\lambda)$ is the corresponding solution of the ZS system~(\ref{ZH system_a})-(\ref{ZH system_b}) at the initial time $t=0$. 
Then, the $n$-SS containing the first $n$ solitons is given by 
\begin{eqnarray}\label{psi_n}
\psi_{(n)}(x,0) = \psi_{(n-1)}(x,0) + 2i(\lambda_n - \lambda^*_n)\frac{q^*_{n1}q_{n2}}{|\mathbf{q_n}|^2},
\end{eqnarray}
where vector $\mathbf{q}_{n}=(q_{n1},q_{n2})^{T}$ is determined by $\mathbf{\Phi}_{(n-1)}$ and the scattering data of the $n$-th soliton $\{\lambda_{n},C_{n}\}$, 
\begin{eqnarray}\label{qn}
\mathbf{q}_{n}(x) = \mathbf{\Phi}^*_{(n-1)}(x,\lambda_n^*)\cdot 
\left(\begin{array}{c} 1 \\C_n \end{array}\right).
\end{eqnarray}
The corresponding solution $\mathbf{\Phi}_{(n)}(x,\lambda)$ of the ZS system is calculated through $\mathbf{\Phi}_{(n-1)}$ and the so-called dressing matrix $\boldsymbol{\chi}$, 
\begin{eqnarray}
\mathbf{\Phi}_{(n)}(x,\lambda) &=& \boldsymbol{\chi}(x,\lambda)\cdot \mathbf{\Phi}_{(n-1)}(x,\lambda), \label{dressing Psi}\\
\boldsymbol{\chi}_{ml}(x,\lambda) &=& \delta_{ml} + \frac{\lambda_n - \lambda^*_n}{\lambda - \lambda_n} \frac{q^*_{nm}q_{nl}}{|\mathbf{q_n}|^2}, \label{dressing matrix}
\end{eqnarray}
where $m,l=1,2$ and $\delta_{ml}$ is the Kronecker delta. 
The recurrent dressing procedure starts from the trivial solution of the NLS equation $\psi_{(0)}=0$ and the corresponding solution of the ZS system 
\begin{eqnarray}\label{Psi0}
\mathbf{\Phi}_{(0)}(x,\lambda) = \begin{pmatrix}\ e^{-i\lambda x} & 0 \\ 0 & e^{i\lambda x} \end{pmatrix},
\end{eqnarray}
and allows one to construct multi-soliton solutions by adding one soliton at each step. 
The dependence in time can be recovered using time-evolution of the scattering data~(\ref{ScattData(t)}) and repeating the dressing method for each moment of time. 

Note that the $N$-SS can also be found via the ratio of two determinants, see e.g.~\cite{novikov1984theory,faddeev2007hamiltonian} and also~\cite{gelash2014superregular}, 
\begin{eqnarray}
\psi_{(N)}(x,t) = 2i\frac{\mathrm{det} \widetilde{M}}{\mathrm{det} M}, \quad
M_{nm}=\frac{(\mathbf{\tilde{q}}_{n}\cdot \mathbf{\tilde{q}}^*_{m})}{\lambda_{n} - \lambda^*_m}, \nonumber\\
\widetilde{M}=
\left(\begin{array}{cc}
        0 & \begin{array}{ccc}
              \tilde{q}_{1,2} & \cdots & \tilde{q}_{N,2}
            \end{array}
         \\
        \begin{array}{c}
          \tilde{q}^*_{1,1} \\
          \vdots \\
          \tilde{q}^*_{N,1}
        \end{array}
         &  \begin{array}{c}
              M^{T}_{nm}
            \end{array}
      \end{array}\right),
\label{Ndet_SS}
\end{eqnarray}
where 
\begin{eqnarray}
\mathbf{\tilde{q}}_{n}(x,t) &=& (\tilde{q}_{n,1},\tilde{q}_{n,2})^{T} = (e^{-\phi_{n}}, e^{\phi_{n}})^{T}, \nonumber\\
\phi_{n}(x,t) &=& -i\lambda_n (x-x_{0n}) - i\lambda_n^2 t - i\theta_n/2.
\end{eqnarray}
However, as we note in the next Section, Eq.~(\ref{Ndet_SS}) is significantly less stable with respect to numerical round-off errors than the described procedure via the dressing method. 


\section{Numerical methods}
\label{Sec:NumMethods}

\subsection{Initial conditions}

With the dressing method described in the previous Section, we generate multi-soliton solutions at the initial time $t=0$ using \textit{Wolfram Mathematica} software with $100$-digits precision. 
This allows us to reliably construct $N$-SS with $N$ more than $100$. 
With the standard $16$-digits precision the dressing method works well up to $N\sim 30$, while the determinant formula~(\ref{Ndet_SS}) fails due to numerical errors already from $N\sim 10$ with the output containing extra large spatial gradients. 
The main source of these errors is the round-off during summation of exponentially small and large values enhanced through the ill-conditioning of the IST problem. 
In addition to better reliability, the dressing method uses $\mathcal{O}(N^2)$ operations to calculate $N$-SS at one space point, while calculation of the determinants demands $\mathcal{O}(N^3)$ operations. 
We think that the difference in the number of operations is the main source of the better numerical stability of the dressing method. 
We did not check performance of the determinant formula with $100$-digits precision, as it takes too much computational time. 

We calculate $N$-SS in the numerical box $x\in[-L/2, L/2]$, limiting soliton positions $x_{0n}$ in such a way that the generated multi-soliton solution is small near the edges of the box.
Specifically, we choose $x_{0n}$ as random values uniformly distributed in a smaller box $x_{0n}\in[-\tilde{L}/2, \tilde{L}/2]$, $\tilde{L}<L$, setting $\tilde{L}$ so that 
$$
|\psi(\pm L/2,0)|\lesssim 10^{-16}\max_{-L/2\le x\le L/2}|\psi(x,0)|.
$$
Later this allows us to treat the box $L$ as periodic and simulate evolution of $N$-SS inside it. 
The soliton phases are generated as uniformly distributed random values too, in the interval $\theta_{n}\in[0,2\pi)$. 
With our methods, we can reliably construct multi-soliton solutions with randomized soliton amplitudes and velocities, and with soliton density 
\begin{equation}
\rho = \frac{2N}{L \mathcal{A}},\quad \mathcal{A}=\frac{1}{N}\sum_{n}a_{n},
\label{density}
\end{equation}
of up to $0.65$. 
Here $\mathcal{A}$ is the average soliton amplitude; our choice for the definition of soliton density is explained below. 
For larger densities, it is difficult to generate ensembles of initial conditions with random soliton parameters, since some of the realizations have to be skipped due to solitons not fitting into the box $L$. 
The latter is the result of the presence of close soliton eigenvalues when solitons interact remotely except for the case of a particular phase synchronization, see e.g.~\cite{gelash2014superregular}. 

For $N$-SS, we use Gaussian-distributed soliton velocities with zero mean and standard deviation $V_{0}$, $v_{n}\sim\mathcal{N}(0, V_{0}^{2})$; we examined the uniform distribution of velocities as well and didn't find essential difference. 
For $a_{n}$, we mainly study the case of equal amplitudes $a_{n}=A=\pi/3.2$, briefly demonstrating the results for Gaussian-distributed amplitudes for comparison. 
Note that in the latter case, for each realization within the ensemble, we first generate amplitudes $a_{n}$ and then shift them by a constant $\delta a$ in order to fix the average intensity 
\begin{equation}
\overline{|\psi|^{2}} = \frac{1}{L}\int_{-L/2}^{L/2}|\psi|^{2}\,dx = \frac{2}{L}\sum_{k=1}^{N}a_{n}
\label{integral_waveaction}
\end{equation}
to the same constant for all initial conditions, see Eqs.~(\ref{integrals_rec}),~(\ref{integrals}) below; otherwise, averaging across ensemble wouldn't be representative. 
The value of $A\approx 0.98$ is chosen to allow direct comparison with the study~\cite{agafontsev2016integrable} of integrable turbulence generated from modulational instability of cnoidal waves. 
After initial generation of soliton velocities $v_{n}$, for each initial condition we shift all $v_{n}$ by a constant $\delta v$ to make the momentum equal to zero, 
\begin{equation}
P = \frac{i}{2L}\int_{-L/2}^{L/2}(\psi_{x}^{*}\psi-\psi_{x}\psi^{*})\,dx = \frac{2}{L}\sum_{n=1}^N a_n v_n = 0. \label{integral_momentum}
\end{equation}

Relation~(\ref{integral_waveaction}) explains our definition of soliton density~(\ref{density}), as in this case the density equals to the ratio between the average square amplitude of the wavefield and square average soliton amplitude, 
$$
\rho=\frac{\overline{|\psi|^{2}}}{\mathcal{A}^{2}}.
$$
Thus, the case $\rho\ll 1$ corresponds to weak interactions between the solitons within rarefied $N$-SS, while $\rho\sim 1$ describes sufficiently dense multi-soliton solutions. 

\begin{figure}[t]\centering
\includegraphics[width=8.5cm]{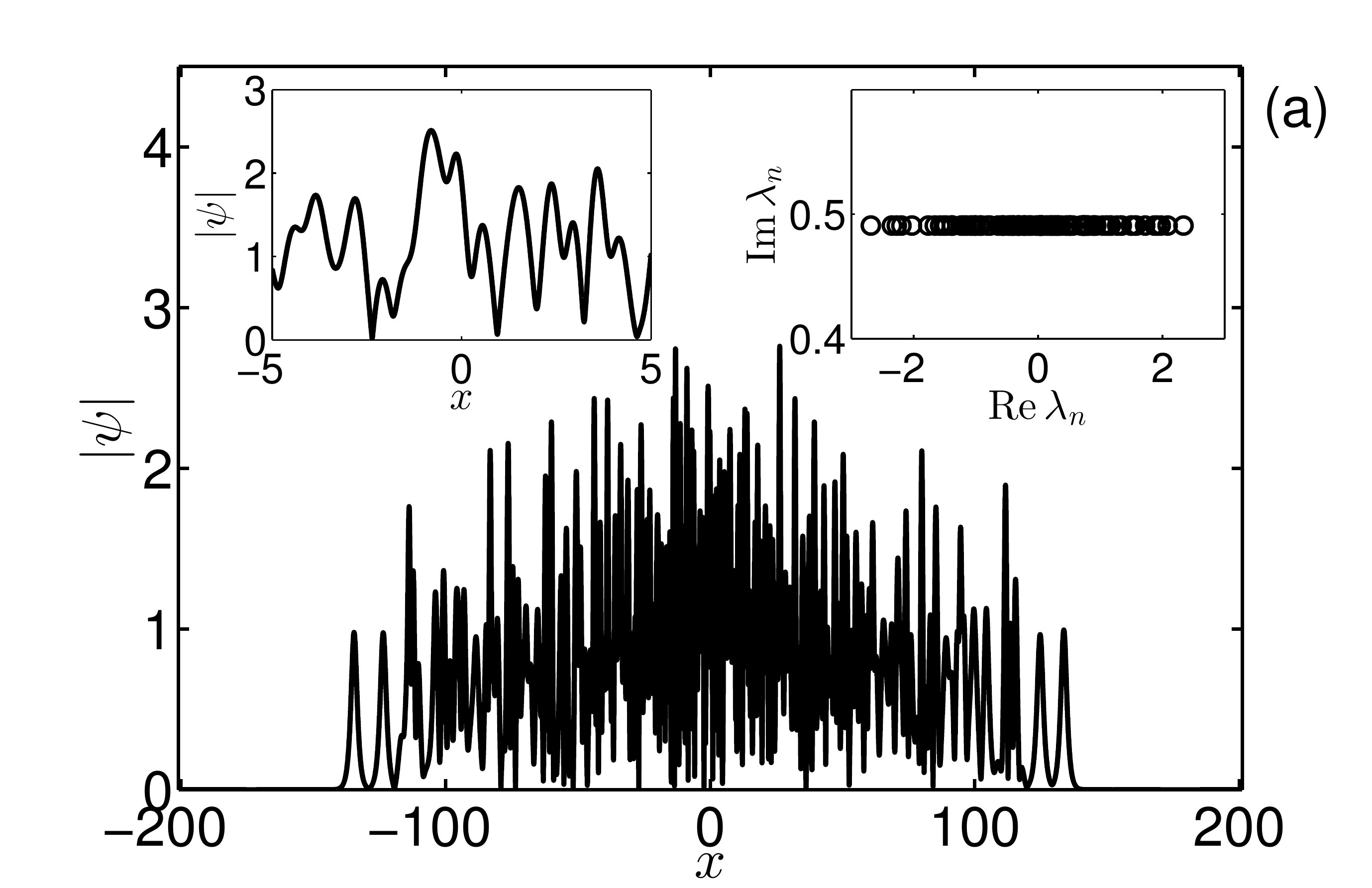}\\
\includegraphics[width=8.5cm]{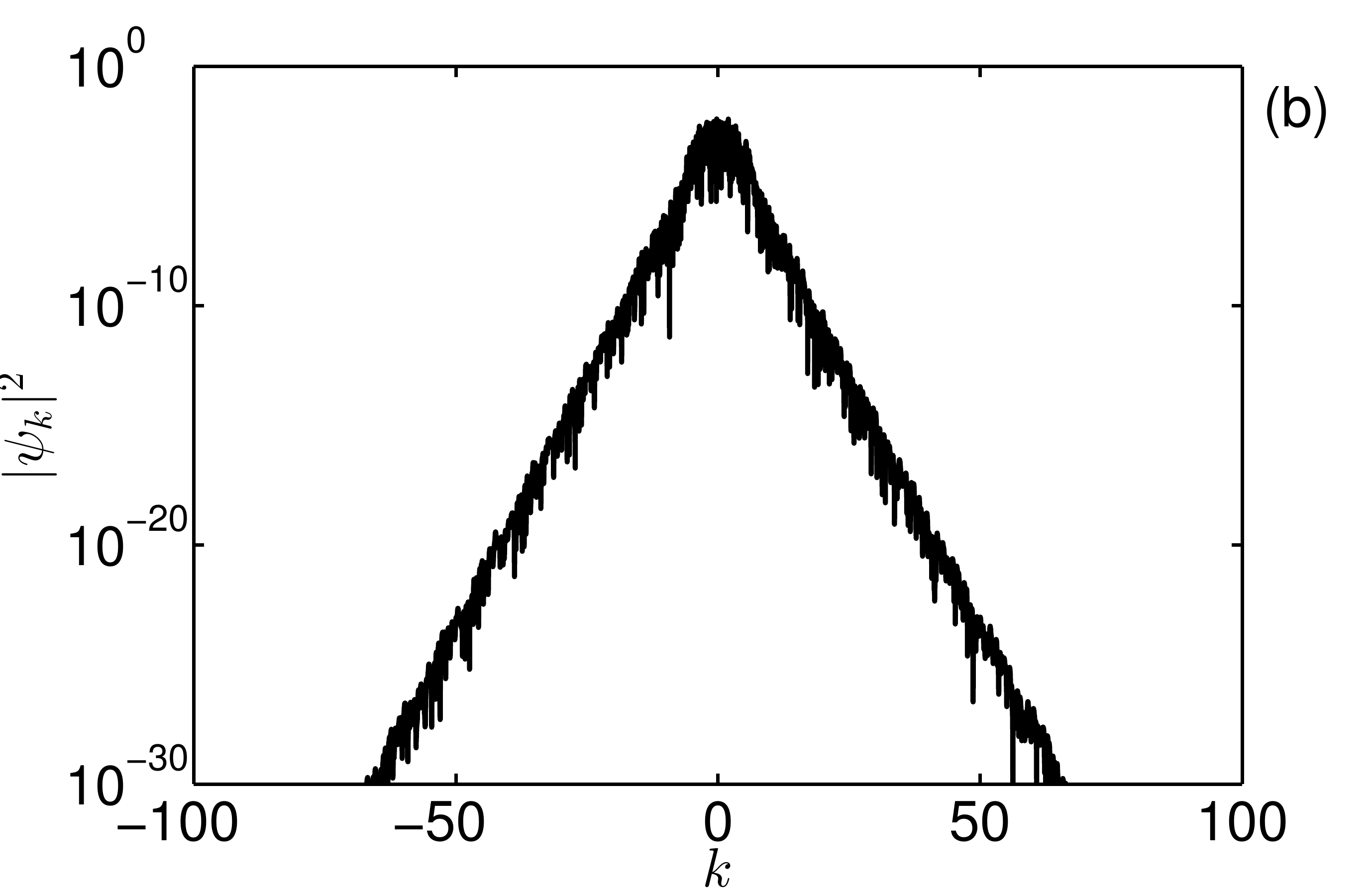}\\
\includegraphics[width=8.5cm]{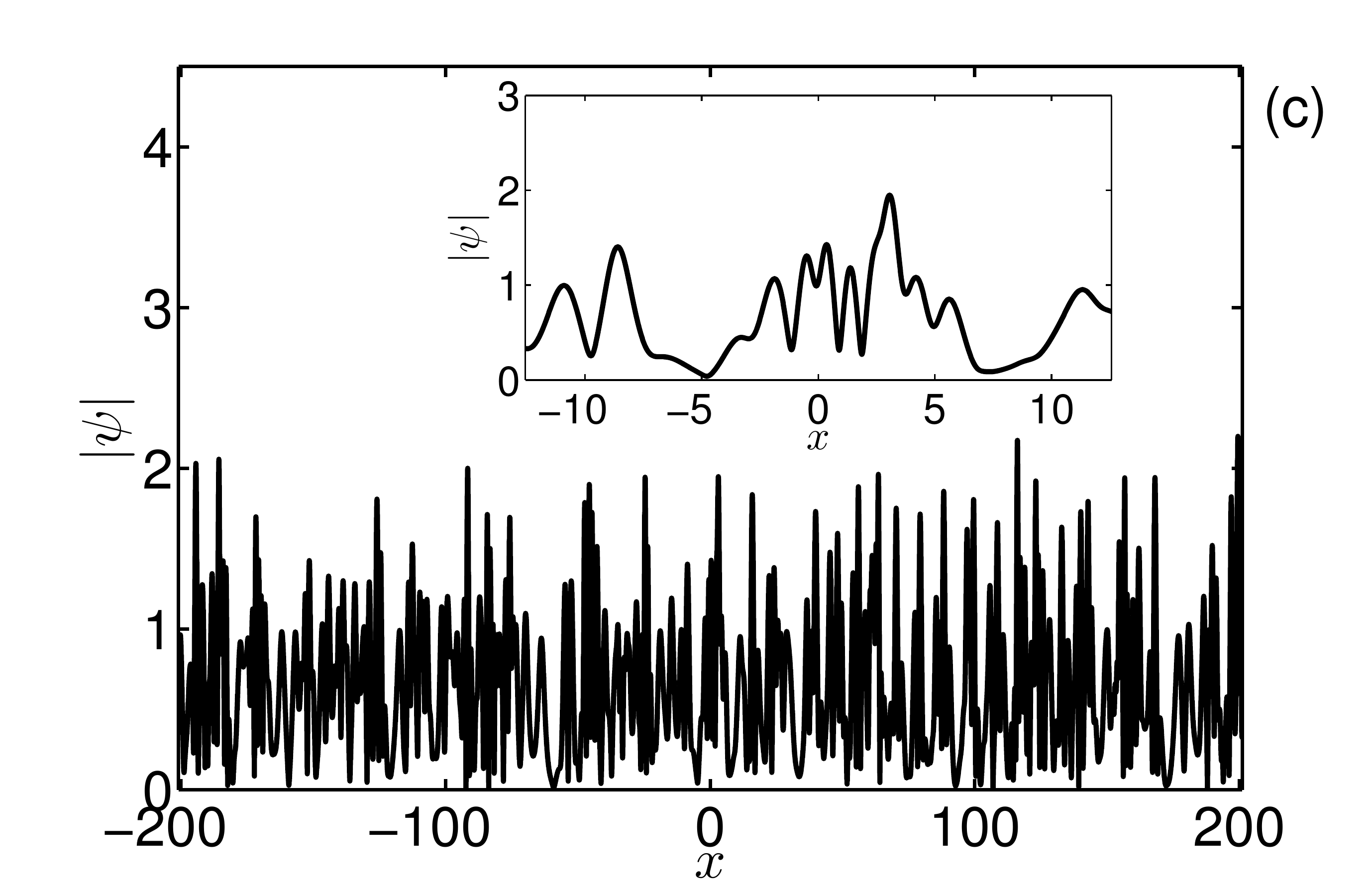}

\caption{\small {\it 
(a) The wavefield of $128$-SS at the initial time $t=0$, (b) its Fourier spectrum $|\psi_{k}|^{2}$ and (c) the wavefield for the same $128$-SS at $t=200$. The simulation box $x\in[-L/2, L/2]$ has length $L=128\pi$, the soliton density is $\rho\approx 0.65$, the amplitudes are equal, $a_{n}=A=\pi/3.2$, and the velocities are Gaussian-distributed, $v_{n}\sim\mathcal{N}(0, V_{0}^{2})$, $V_{0}=2$. 
The left inset in figure (a) and the inset in figure (c) illustrate the wavefield at the center of the box $L$ with better resolution, and the right inset in figure (a) shows the eigenvalues $\lambda_{n}$, $n=1,...,128$. 
}}
\label{fig:fig01}
\end{figure}

If two eigenvalues $\lambda_{n}$ and $\lambda_{m}$ turn out to be close to each other, the numerical implementation of both the dressing method and the determinant formula may fail due to very small denominators in Eqs.~(\ref{dressing matrix}),~(\ref{Ndet_SS}) (for $\lambda_{n}=\lambda_{m}$ these equations become indeterminate). 
To avoid such situations, we use threshold $\delta\lambda=10^{-9}$ for minimal distance between the eigenvalues 
\begin{equation}
|\lambda_n - \lambda_m | > \delta\lambda, \quad n,m=1,...,N, \quad m\neq n. \label{lambda_threshold}
\end{equation}
Technically, we generate $\lambda_{n}$ in sequential order, check for $\lambda_{n}$ if the relation~(\ref{lambda_threshold}) is valid for all $m<n$, and if it doesn't, we generate $\lambda_{n}$ again, for $n=2,...,N$. 
For our velocity distributions, the probability of one event when any $\lambda_{n}$ has to be recalculated is of $10^{-5}$ order. 
This is why we think that the influence of this threshold on the statistical characteristics of soliton gas can be neglected. 

An $128$-soliton solution generated with the above described rules together with its Fourier spectrum is shown in Fig.~\ref{fig:fig01}(a,b).

\subsection{Time evolution}

We generate ensembles containing $\sim 10^{3}$ individual realizations of $N$-SS (we use ensembles of $1000$ for $128$-SS and $2000$ for $64$-SS). 
Then, we simulate their evolution solving the NLS equation~(\ref{NLSE}) numerically and treating the box $x\in[-L/2, L/2]$ where the $N$-SS were generated as periodic, so that with time the solitons spread over the box (statistically) uniformly. 
With this approach, we actually change $N$-SS by $N$-gap solutions with very narrow gaps~\cite{belokolos1994algebro,bobenko2011computational} (soliton widths are by two-three orders of magnitude smaller than the box size), neglecting the difference between the two.  
We use pseudospectral Runge-Kutta 4th-order method with adaptive change of the spatial grid size $\Delta x$ and Fourier interpolation between the grids, as described in~\cite{agafontsev2015integrable,agafontsev2016integrable}; $\Delta x$ is set from the analysis of the Fourier spectrum of the solution. 
The time step $\Delta t$ changes with $\Delta x$ as $\Delta t = h\,\Delta x^{2}$, $h\le 0.1$, in order to avoid numerical instabilities. 

Integrability of the NLS equation implies conservation of infinite series of integrals of motion~\cite{novikov1984theory}, 
\begin{eqnarray}
\mathcal{I}_{n} &=& \frac{1}{L}\int_{-L/2}^{L/2}f_{n}\,dx, \nonumber\\
f_{n} &=& \psi\frac{\partial}{\partial x}\bigg(\frac{f_{n-1}}{\psi}\bigg)+\sum_{l+m=n-1}f_{l}f_{m},
\label{integrals_rec}
\end{eqnarray}
where $f_{1}=|\psi|^{2}$. 
The first three of these invariants are wave action (in our notations equals to average intensity)
\begin{equation}\label{wave-action}
S = \overline{|\psi|^{2}} = \frac{1}{L}\int_{-L/2}^{L/2}|\psi|^{2}\,dx,
\end{equation}
momentum~(\ref{integral_momentum}) and total energy 
\begin{eqnarray}\label{energy}
&& E = H_{d} + H_{4},\quad H_{d}=\frac{1}{L}\int_{-L/2}^{L/2}|\psi_{x}|^{2}\,dx,\nonumber\\
&& H_{4} = -\frac{1}{L}\int_{-L/2}^{L/2}|\psi|^{4}\,dx. 
\end{eqnarray}
Here $H_{d}$ is the kinetic energy and $H_{4}$ is the potential one, and we use prefactor $1/L$ for further convenience. 
For $N$-SS, these invariants can also be found via the eigenvalues, 
\begin{equation}\label{integrals}
\mathcal{I}_{n} = \frac{(2i)^n}{nL}\sum_{k=1}^N [(\lambda_k^n)^* - \lambda_k^n].
\end{equation}
In our simulations, the first ten integrals calculated via the recurrent formula~(\ref{integrals_rec}) conserve with time and coincide with the exact values~(\ref{integrals}) with the relative errors from $10^{-14}$ (for $\mathcal{I}_{1}$) to $10^{-8}$ (for $\mathcal{I}_{10}$) orders. 
Additionally, we compare discrete eigenvalues $\lambda_{n}$ of the initial conditions with eigenvalues $\Lambda_{n}$ calculated at the final time of the evolution, which we find by solving the ZS eigenvalue problem~(\ref{ZH system1}) with the Fourier collocation method, see e.g~\cite{yang2010nonlinear,randoux2016inverse}; the eigenvalues coincide up to the relative error $|\lambda_{n}-\Lambda_{n}|/|\lambda_{n}|$ of $10^{-9}$ order. 

Due to periodic border conditions, the solitons spread over the computational box sufficiently uniformly to time $t\sim 80$ for our $N$-SS parameters, and the system arrives to its statistically steady state, where its ensemble-averaged characteristics do not change anymore. 
In particular, among these characteristics we check the moments of amplitude $M_{n}=\langle\overline{|\psi|^{n}}\rangle$ (using them to determine the steady state), the wave-action spectrum
\begin{equation}\label{wave-action-spectrum}
S_{k} = \langle|\psi_{k}|^{2}\rangle,\quad \psi_{k} = \frac{1}{L}\int_{-L/2}^{L/2}\psi\, e^{-ikx}\,dx,
\end{equation}
the kinetic $\langle H_{d}\rangle$ and potential $\langle H_{4}\rangle$ energies, the kurtosis $\kappa=M_{4}/M_{2}^{2}$, and the PDF of relative wave intensity, see e.g.~\cite{agafontsev2015integrable,agafontsev2016integrable}. 
Here $\langle ...\rangle$ means arithmetic averaging across ensemble of initial conditions, while the overline denotes spatial averaging, see e.g.~(\ref{wave-action}). 
We use the achieved statistically steady state as a model of homogeneous soliton gas of density~(\ref{density}) in an infinite space; we confirmed that for $N\gtrsim 32$ and $L\gtrsim 32\pi$ the statistical properties of the steady state depend on $N$ and $L$ only in combination $N/L\propto\rho$. 
We perform simulations until $t=200$ and additionally average the results in the time interval $t\in[160,200]$; an example of the wavefield at $t=200$ is shown in Fig.~\ref{fig:fig01}(c). 
For smaller soliton velocities we shift to larger times, which we determine according to the same principle as described above. 

As we noted earlier, see Eqs.~(\ref{integral_waveaction})-(\ref{integral_momentum}), all realizations within the ensemble of initial conditions have zero momentum $P=0$ and the same value of wave action (average intensity) $S=\overline{|\psi|^{2}}$. 
However, as one can see from Eq.~(\ref{integrals}), the integrals of motion of higher order are not fixed and may change significantly from one realization to another. 
Of these integrals, the total energy can be fixed by fixing the average square velocity $u^{2}=(1/N)\sum_{n} v_{n}^{2}$ for each realization, see Eq.~(\ref{raregas}) for the total energy below. 
Technically, this can be done by multiplying each soliton velocity $v_{n}$ by a constant to make $u^{2}$ the same for each realization. 
We compared our results against those obtained using $N$-SS with the fixed value of the total energy and found no difference. 

We checked our statistical results against the size of the ensembles and parameters of our numerical scheme, and found no difference. 
We also found that, for random soliton velocities, the statistical properties of the steady state do not depend on distributions of positions $x_{0n}$ and complex phases $\theta_{n}$. 
The latter result is straightforward from the point of view of soliton collisions. 
Indeed, when solitons collide, they experience jumps in positions and complex phases, which depend on the velocity difference~\cite{novikov1984theory}. 
If soliton velocities are random, these jumps lead to stohastization of positions and complex phases even when their initial values are not random. 
Thus, for random velocities, the statistical characteristics of homogeneous soliton gas depend only on soliton density and distributions of amplitudes and velocities. 


\section{Results}
\label{Sec:Results}

In the limit of small soliton density, soliton gas is a superposition of almost non-interacting individual solitons moving with different velocities, and the wavefield is very close to arithmetic sum of one-soliton solutions~(\ref{1-SS}). 
Then, neglecting soliton interactions and using Eqs.~(\ref{integral_waveaction}),~(\ref{energy}) and definition of soliton density~(\ref{density}), one can find the average intensity, the kinetic, potential and total energies, and the kurtosis, respectively 
\begin{eqnarray}
&& \langle\overline{|\psi|^{2}}\rangle = A^{2}\rho, \quad \langle H_{d}\rangle = \frac{A^{2}\rho}{3}(3u^{2}+A^{2}),\nonumber\\
&& \langle H_{4}\rangle = -\frac{2A^{4}\rho}{3}, \quad \langle E\rangle = \frac{A^{2}\rho}{3}(3u^{2}-A^{2}), \nonumber\\
&& \kappa = \frac{\langle\overline{|\psi|^{4}}\rangle}{\langle\overline{|\psi|^{2}}\rangle^{2}} = \frac{2}{3\rho}. \label{raregas}
\end{eqnarray}
Here $u^{2}=\langle v_{n}^{2}\rangle$ is the average square velocity, and we consider soliton gas with equal amplitudes $a_{n}=A$ for simplicity. 
Relations for average intensity and total energy are exact ones and follow directly from Eqs.~(\ref{integral_waveaction}),~(\ref{integrals}), while the other ones for the kinetic energy, the potential energy and the kurtosis are approximate and valid only when $\rho\ll 1$. 

\begin{figure}[t]\centering
\includegraphics[width=8.5cm]{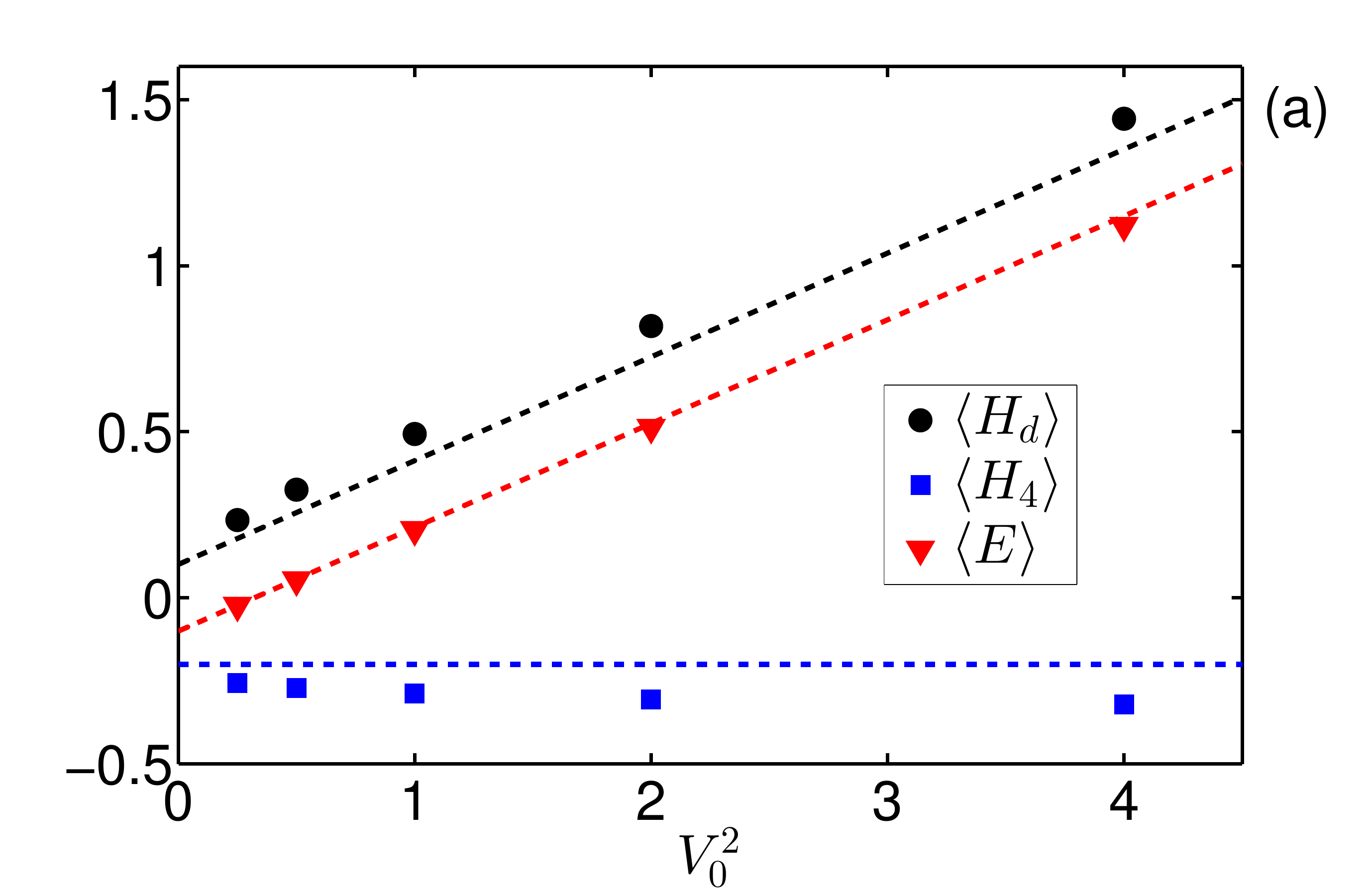}\\
\includegraphics[width=8.5cm]{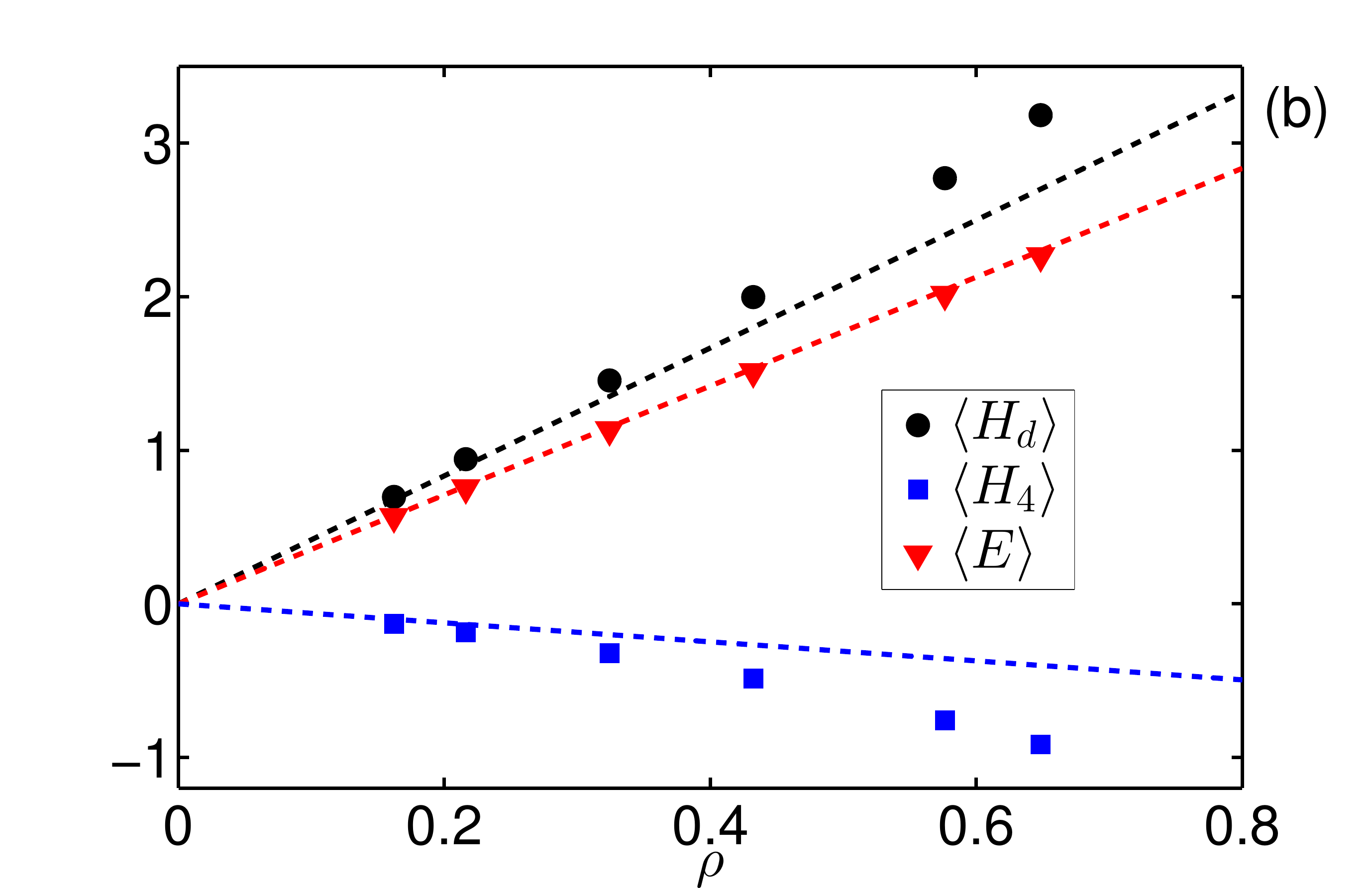}

\caption{\small {\it (Color on-line) 
(a) The kinetic $\langle H_{d}\rangle$, potential $\langle H_{4}\rangle$ and total $\langle E\rangle$ energies for statistically homogeneous soliton gas with density $\rho=0.32$ and equal amplitudes $a_{n}=A=\pi/3.2$, depending on the average square velocity $V_{0}^{2}$; we model these gases with $64$-SS. 
The dashed lines indicate the relations~(\ref{raregas}) with $u^{2}=V_{0}^{2}$ for the kinetic (black), potential (blue) and total (red) energies, respectively. 
(b) The same for soliton gas with fixed characteristic velocity $V_{0}=2$, depending on soliton density $\rho$; we model these gases with $128$-SS. 
}}
\label{fig:fig02}
\end{figure}

\begin{figure}[t]\centering
\includegraphics[width=8.5cm]{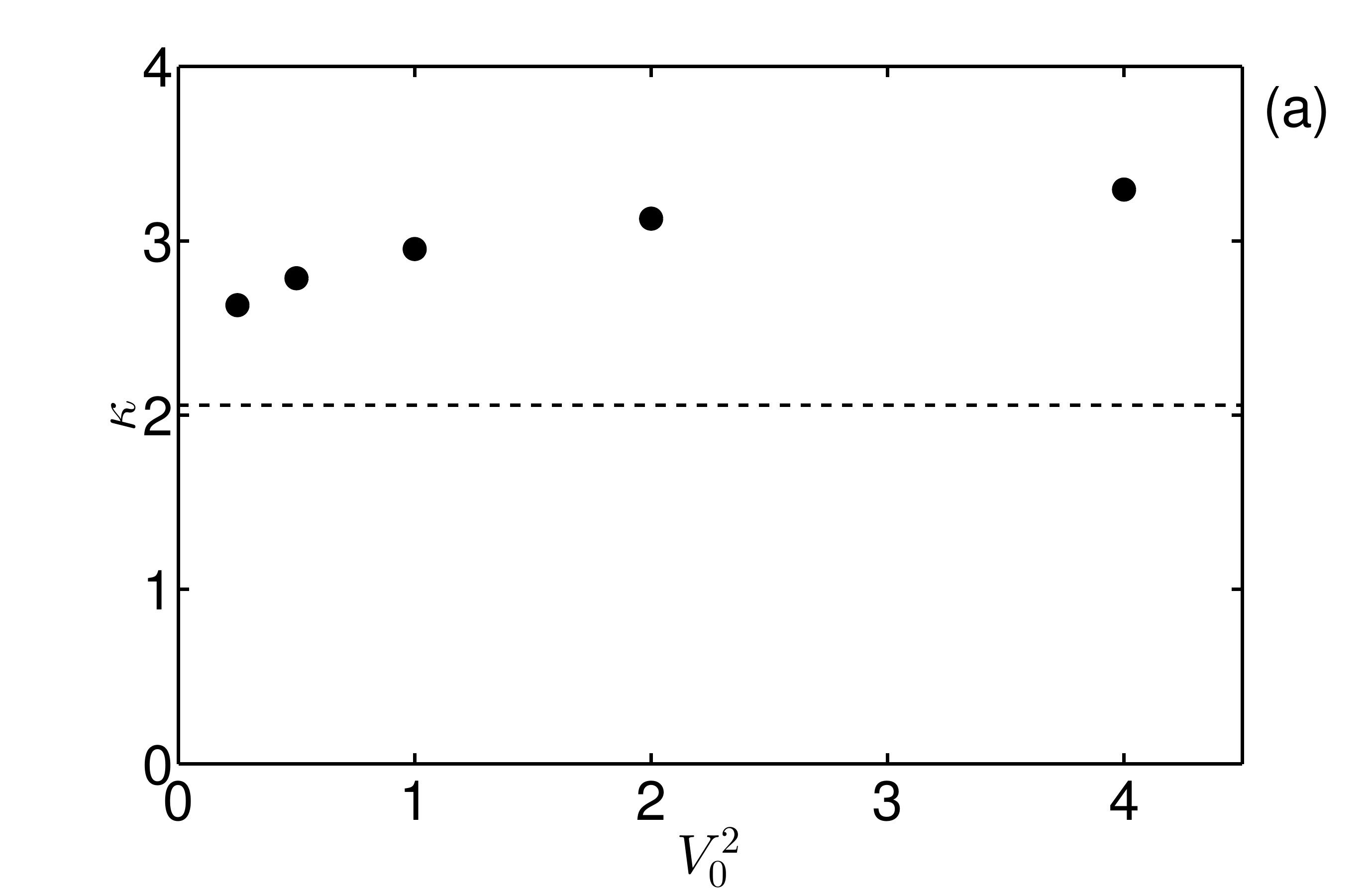}\\
\includegraphics[width=8.5cm]{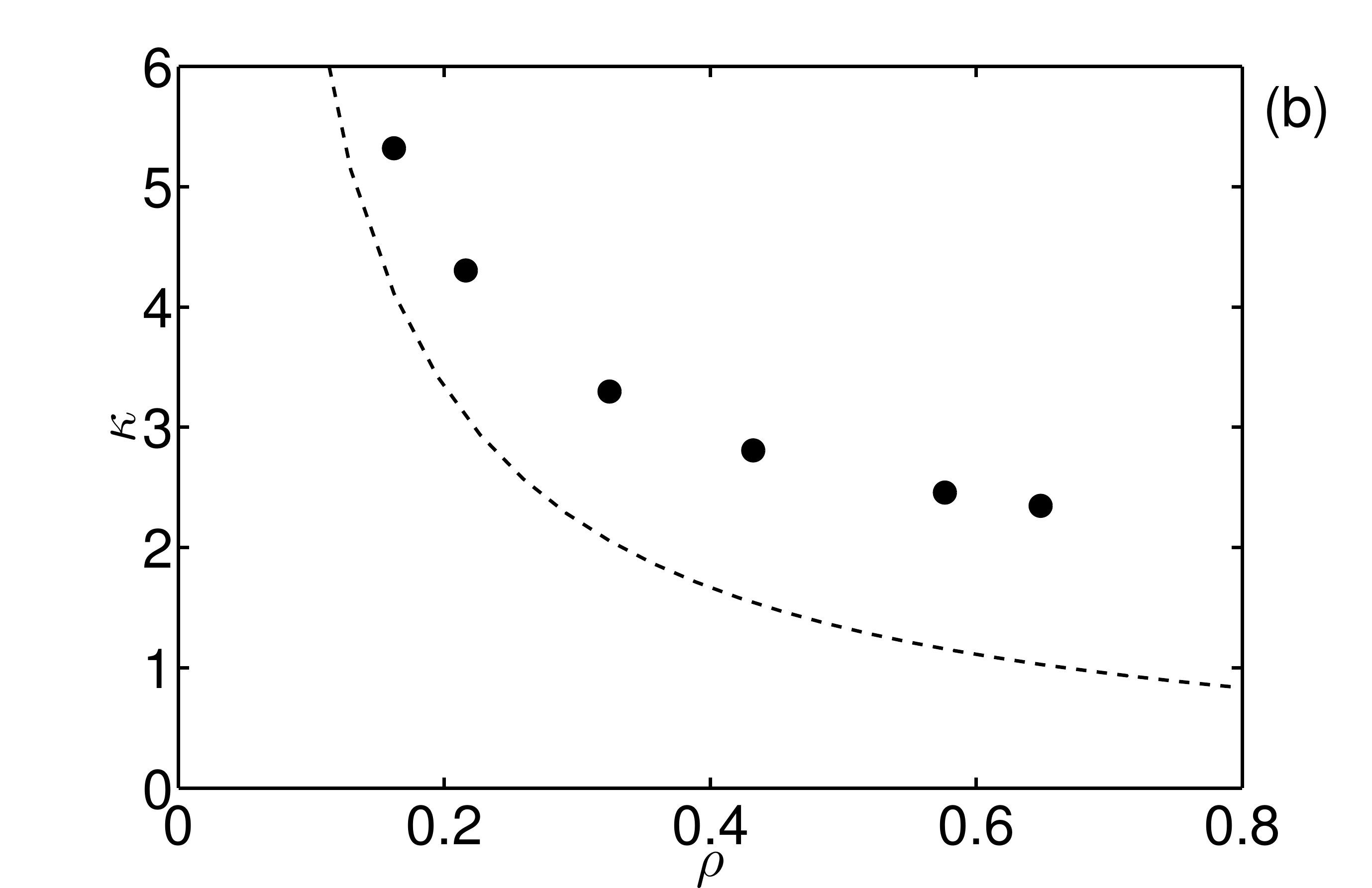}

\caption{\small {\it 
(a) The kurtosis $\kappa$ depending on the average square velocity $V_{0}^{2}$; all the parameters are the same as in Fig.~\ref{fig:fig02}(a). 
The dashed line indicates the relation~(\ref{raregas}) for the kurtosis, with $u^{2}=V_{0}^{2}$. 
(b) The kurtosis $\kappa$ depending on the soliton density $\rho$; all the parameters are the same as in Fig.~\ref{fig:fig02}(b).
}}
\label{fig:fig03}
\end{figure}

To test these relations, we perform simulations with several ensembles of initial conditions, which differ from each other by characteristic soliton velocity $V_{0}$ or soliton density $\rho$. 
Using these ensembles we model statistically homogeneous soliton gas as described in the previous Section, finding numerically the values for the kinetic, potential and total energies, as well as the kurtosis; these results are shown in Fig.~\ref{fig:fig02} and~\ref{fig:fig03} depending on $V_{0}^{2}$ and $\rho$. 
Note that, as we use Gaussian distribution for the velocities $v_{n}\sim\mathcal{N}(0, V_{0}^{2})$, the average square velocity coincides with the variance, $u^{2}=V_{0}^{2}$. 
Numerical results correspond very well with the relations~(\ref{raregas}). 
The deviations represent next-order corrections due to soliton interactions and increase with characteristic velocity $V_{0}$ or density $\rho$. 
Indeed, larger values for $V_{0}$ or $\rho$ lead to enhanced soliton interactions due to more frequent collisions or decreased spacing. 
This results in more frequent appearance of large amplitudes $|\psi|$ and gradients $|\psi_{x}|$ for the wavefield, i.e. in larger absolute values for the potential and kinetic energies, respectively. 
More frequent appearance of large ``spikes'' should increase the kurtosis as well, with respect to its collisionless value. 
We observe this behavior in Fig.~\ref{fig:fig02} and~\ref{fig:fig03}, where the deviations from relations~(\ref{raregas}) increase with $V_{0}$ and $\rho$, and lead to increased absolute values for the kinetic and potential energies, and also the kurtosis. 
Note that the kurtosis $\kappa$ in our simulations approaches to $2$ with increasing density $\rho$, what hints to the possibility that in the limit of dense soliton gas the PDF of relative wave intensity may converge to the exponential PDF~(\ref{Rayleigh}). 

\begin{figure}[t]\centering
\includegraphics[width=8.5cm]{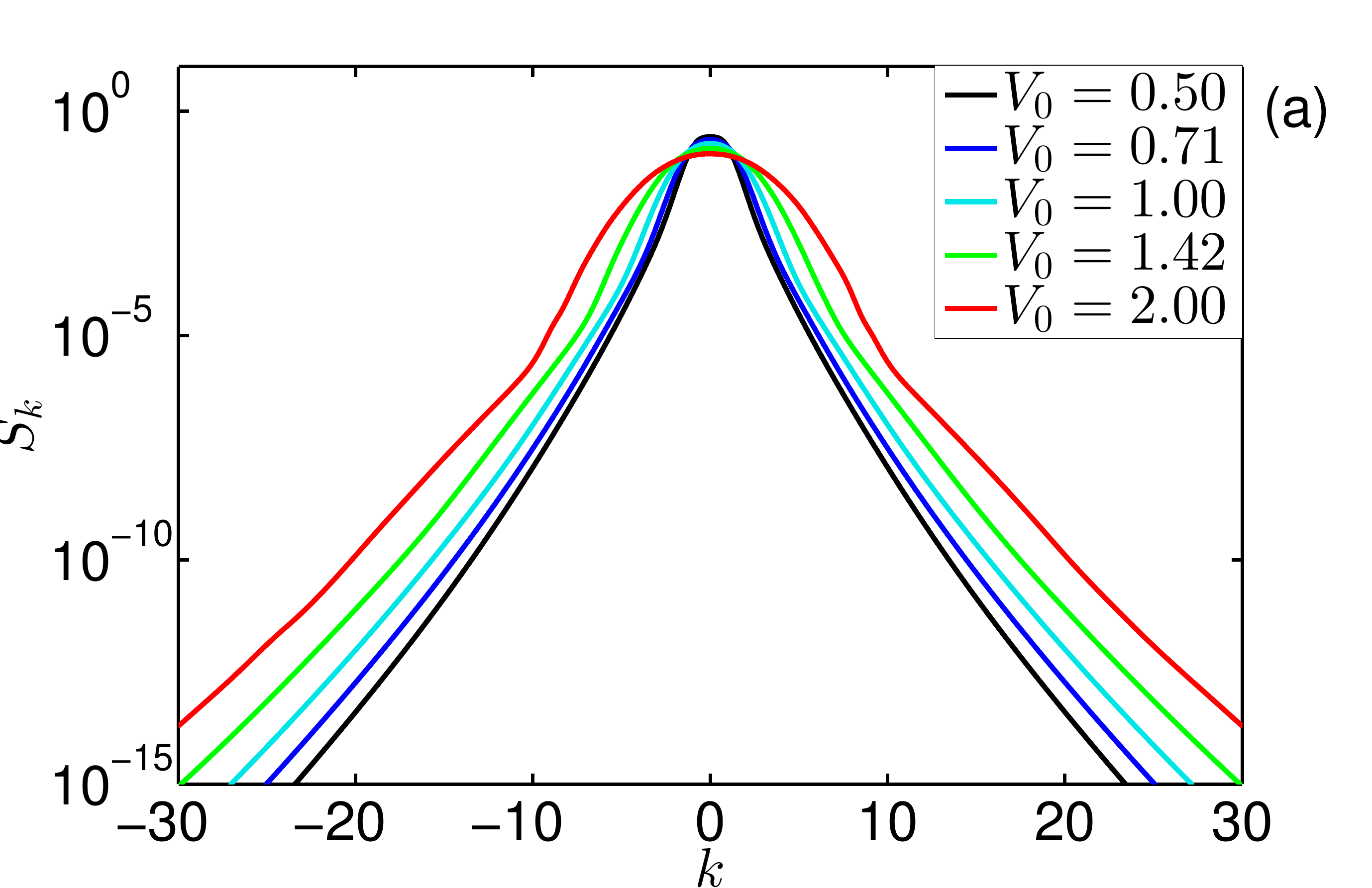}\\
\includegraphics[width=8.5cm]{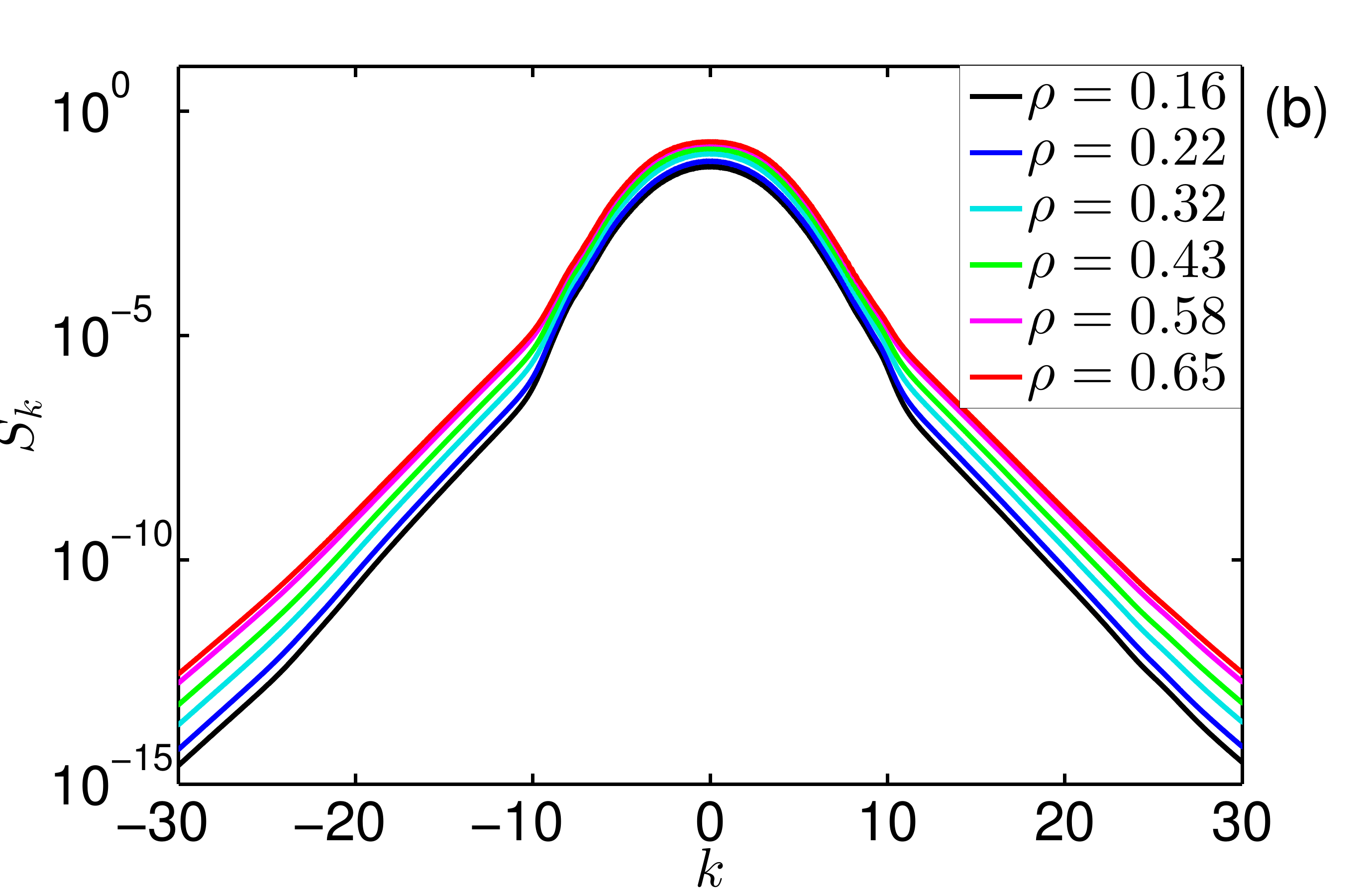}

\caption{\small {\it (Color on-line) 
(a) The wave-action spectrum $S_{k}$ for statistically homogeneous soliton gas, for different characteristic soliton velocities $V_{0}$. All the parameters are the same as in Fig.~\ref{fig:fig02}(a). 
(b) The same for different soliton densities $\rho$; all the parameters are the same as in Fig.~\ref{fig:fig02}(b). 
}}
\label{fig:fig04}
\end{figure}

Fig.~\ref{fig:fig04} shows the wave-action spectrum for different characteristic soliton velocities $V_{0}$ and densities $\rho$. 
In contrast to the condensate~\cite{agafontsev2015integrable} and the cnoidal wave~\cite{agafontsev2016integrable} initial conditions, the spectrum is smooth -- it does not contain peaks diverging by power-law with the wavenumber. 
At small and moderate wavenumbers the spectrum has a characteristic bell-shaped form which significantly depends on the characteristic velocity and density, while at large $k$ the spectrum decays close to exponentially. 
Enhanced soliton interactions with increasing $V_{0}$ and $\rho$ are reflected in widening of the spectrum. 
Note that since in our notations 
$$
\sum_{k}S_{k} = \langle\overline{|\psi|^{2}}\rangle = \mathcal{A}\,\rho,
$$
the sum of the spectrum over the wavenumber depends linearly on the soliton density. 

\begin{figure}[t]\centering
\includegraphics[width=8.5cm]{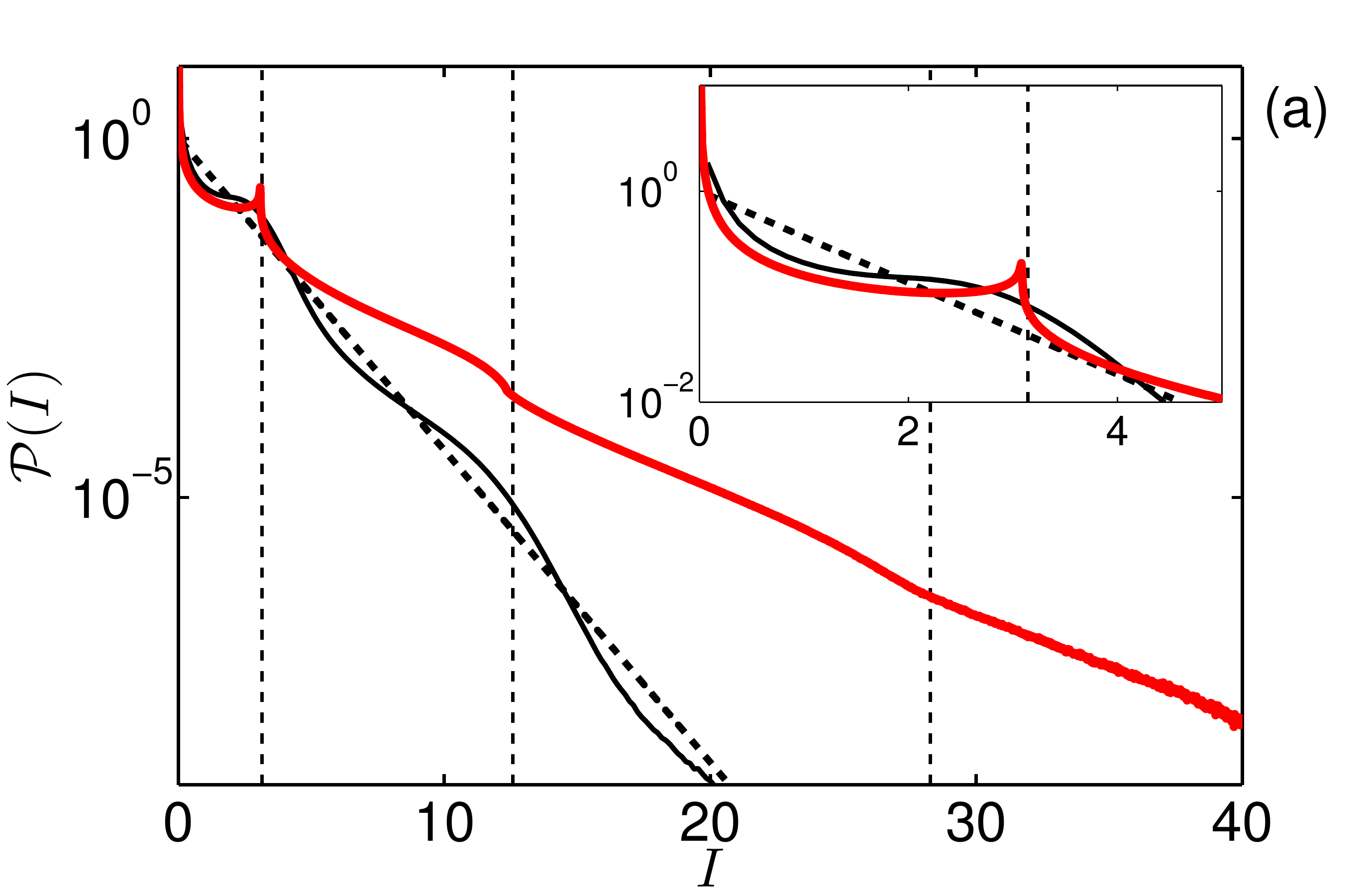}\\
\includegraphics[width=8.5cm]{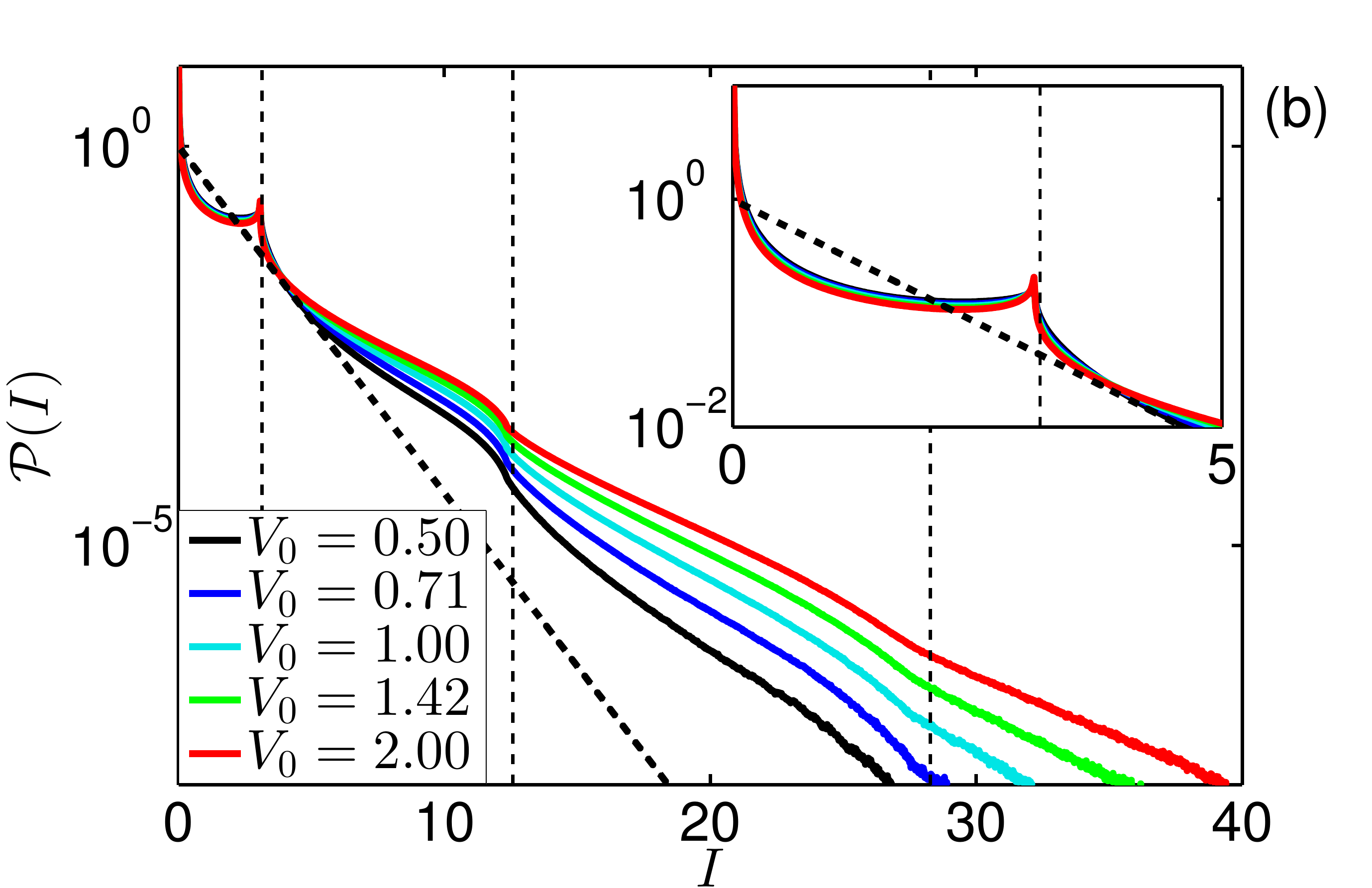}

\caption{\small {\it (Color on-line) 
(a) The PDF $\mathcal{P}(I)$ for statistically homogeneous soliton gas (thick red) and the cnoidal wave initial conditions (thin black), in comparison with the exponential PDF $\mathcal{P}_{R}(I)$ (dashed black). 
The soliton gas has density $\rho=0.32$, equal amplitudes $a_{n}=A=\pi/3.2$ and characteristic velocity $V_{0}=2$; we model it with $64$-SS. 
The cnoidal wave is ``constructed'' from solitons of amplitude $A$ and has the same ``soliton'' density as the soliton gas, what corresponds to $\omega_{1}=1.6$ in notations of~\cite{agafontsev2016integrable}; the PDF is taken from this publication. 
(b) The PDF $\mathcal{P}(I)$ for soliton gas depending on characteristic soliton velocity $V_{0}$; all other parameters are the same as in figure (a). The dashed line shows the exponential PDF $\mathcal{P}_{R}(I)$. 
The thin vertical lines in both figures mark relative intensities $I$ corresponding to amplitudes $|\psi|=A$, $2A$ and $3A$, while the insets show PDFs at small and moderate $I$. 
}}
\label{fig:fig05}
\end{figure}

We examine the PDF starting from the example of soliton gas with density $\rho=0.32$, equal amplitudes $a_{n}=A=\pi/3.2$, and characteristic soliton velocity $V_{0}=2$; we model this gas with $64$-SS. 
The corresponding PDF is shown in Fig.~\ref{fig:fig05}(a), in comparison with the exponential PDF~(\ref{Rayleigh}) and the asymptotic PDF for the cnoidal wave initial conditions, with the cnoidal wave ``constructed'' from solitons of the same amplitude ($\omega_{1}=1.6$ in notations of~\cite{agafontsev2016integrable}). 

Cnoidal waves are exact periodic solutions of the NLS equation which can be represented as lattices of overlapping solitons~\cite{kuznetsov1999modulation}. 
The case $\omega_{1}=1.6$ corresponds to a lattice of solitons with amplitude $A=\pi/3.2$ and with ``soliton'' density $\rho=0.32$ -- the same parameters as for the soliton gas we examine in Fig.~\ref{fig:fig05}(a). 
Modulational instability of such lattices leads to integrable turbulence, which asymptotically approaches to its statistically steady state in an oscillatory way~\cite{agafontsev2016integrable}. 
The PDF in this state depends on the degree of ``overlapping'' of solitons. 
When the overlapping is weak, the wavefield remains close to a collection of thin and high solitons with different phases and positions, soliton amplitude exceeds significantly the space-average amplitude, and the PDF deviates from the exponential distribution $\mathcal{P}_{R}(I)$~(\ref{Rayleigh}) pronouncedly. 
For strong overlapping, the behavior of the system is similar to that of the modulational instability of the condensate~\cite{agafontsev2015integrable} and the PDF in the steady state coincides with $\mathcal{P}_{R}(I)$. 
The case $\omega_{1}=1.6$ shown in Fig.~\ref{fig:fig05}(a) by the thin black line is an ``intermediate'' and mixes properties of these two limits. 
In particular, from one hand the corresponding PDF is relatively close to the exponential, and from the other one it can be subdivided by characteristic parts representing 1) a singular soliton, 2) two-soliton and 3) three-soliton collisions, as demonstrated by the vertical dashed lines in the figure. 
These three lines are drawn at intensities corresponding to amplitudes $A$, $2A$ and $3A$, as the maximal amplitude of $N$-SS equals to $NA$, see e.g.~\cite{akhmediev1991extremely}. 
The same parts can also be distinguished for the PDF of soliton gas in Fig.~\ref{fig:fig05}(a), however, at the soliton interactions region $|\psi|^{2}\ge A^{2}$ this PDF exceeds the exponential distribution $\mathcal{P}_{R}(I)$ by orders of magnitude. 

\begin{figure}[t]\centering
\includegraphics[width=8.5cm]{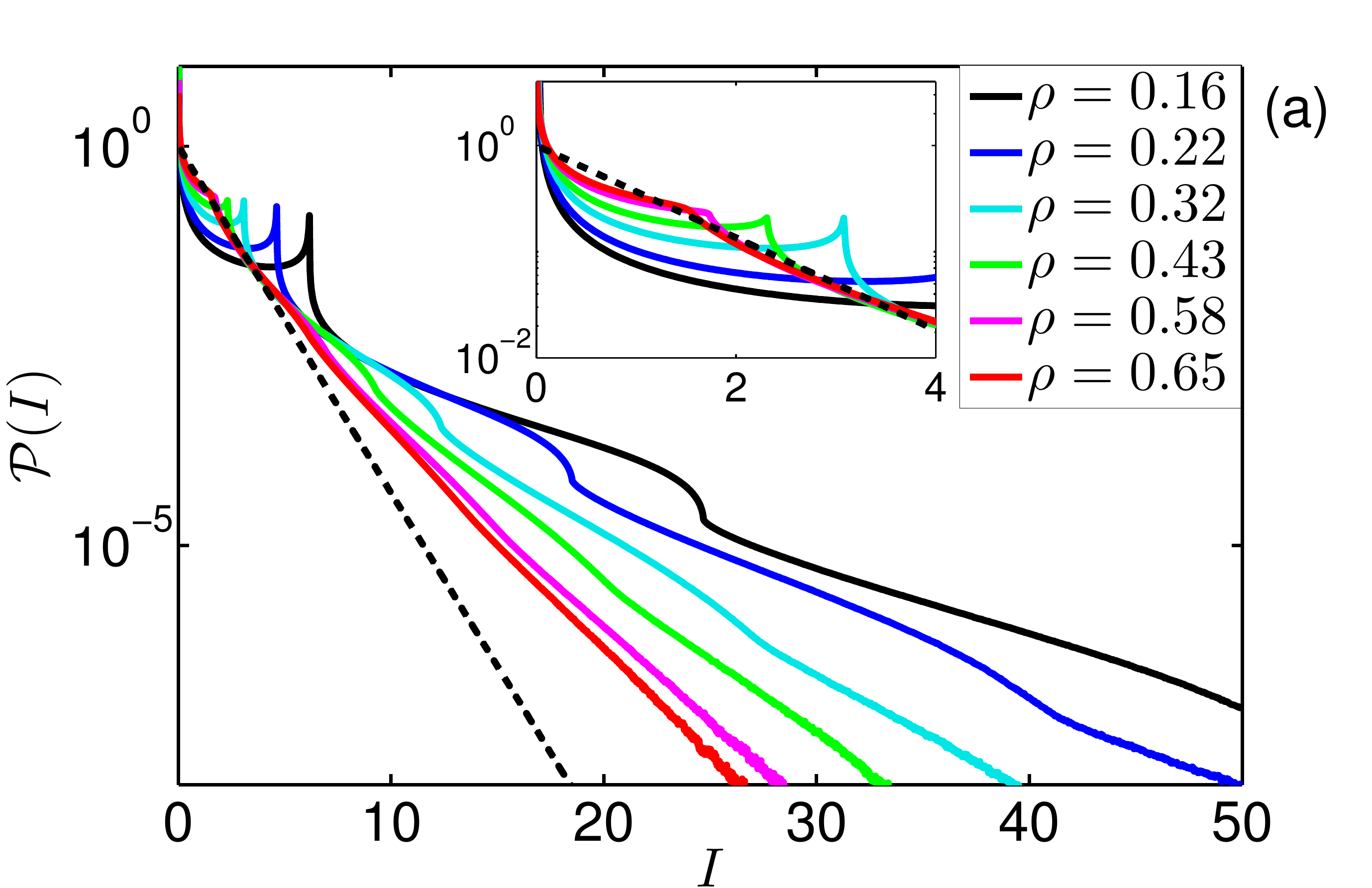}\\
\includegraphics[width=8.5cm]{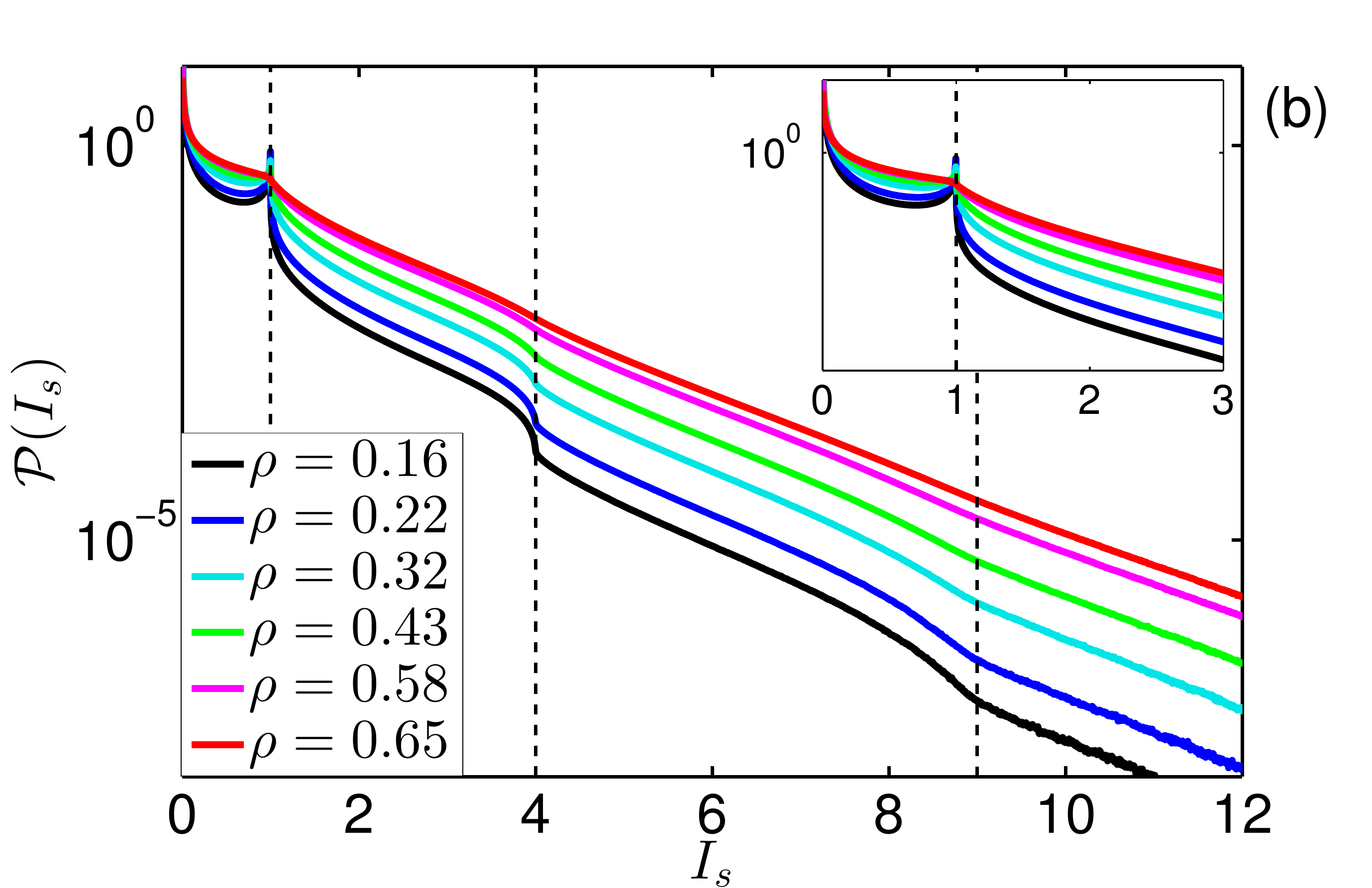}

\caption{\small {\it (Color on-line) 
(a) The PDF $\mathcal{P}(I)$ for statistically homogeneous soliton gas, depending on soliton density $\rho$. 
The soliton gas has equal amplitudes $a_{n}=A=\pi/3.2$ and characteristic velocity $V_{0}=2$; we model it with $128$-SS. 
The dashed line is the exponential PDF $\mathcal{P}_{R}(I)$, the inset shows PDFs at small and moderate relative intensities $I$. 
(b) The same PDFs recalculated for renormalized intensity $I_{s}=|\psi|^{2}/A^{2}$. 
The thin vertical lines correspond to soliton amplitude $A$, and maximal amplitudes of two- $2A$ and three-soliton $3A$ collisions; the inset shows PDFs at small and moderate $I_{s}$. 
}}
\label{fig:fig06}
\end{figure}

As demonstrated in Fig.~\ref{fig:fig05}(b), the excess over $\mathcal{P}_{R}(I)$ is larger for soliton gas with larger velocities.
Note that the cnoidal wave in the limit of small soliton overlapping approaches to soliton gas with zero velocities. 
Thus, we may assume that when the soliton density is small the PDF for soliton gas with small velocities should be close to the asymptotic PDF for the cnoidal wave initial conditions (if the corresponding cnoidal wave with the same soliton density and amplitudes can be constructed). 
However, we cannot reliably model soliton gas with $V_{0}\lesssim 0.5$ to check it, since in this case some of the realizations of $N$-SS do not fit into the computational box and have to be skipped from the ensemble of initial conditions. 

Fig.~\ref{fig:fig06}(a) shows PDF for soliton gas with fixed amplitudes $a_{n}=A$ and characteristic velocity $V_{0}=2$, depending on soliton density $\rho$; we model these gases with $128$-SS. 
When the density is small $\rho\ll 1$, soliton gas is a collection of weakly interacting thin (compared to average spacing) and high (compared to average amplitude $[\overline{|\psi|^{2}}]^{1/2}\ll A$) solitons. 
In this case, even an individual pulse with amplitude close to $A$ may already be a rogue wave, and the PDF deviates drastically from the exponential distribution $\mathcal{P}_{R}(I)$, exceeding it by orders of magnitude at sufficiently large relative intensities $I$. 
The regions on the PDF representing a singular soliton, and two- and three-soliton collisions are clearly visible in Fig.~\ref{fig:fig06}(b), which shows PDFs for renormalized intensity $I_{s}=|\psi|^{2}/A^{2}$; these regions correspond to segments $I_{s}\in[0,1]$, $I_{s}\in[1,4]$ and $I_{s}\in[4,9]$, respectively. 

For larger densities, the solitons collide more frequently, and the PDF $\mathcal{P}(I)$ of relative wave intensity transforms closer to the exponential PDF $\mathcal{P}_{R}(I)$. 
When the density becomes of unity order, $\rho\sim 1$, the collisions become so frequent that the regions on the PDF corresponding to two- and three-soliton collisions are difficult to distinguish, Fig.~\ref{fig:fig06}(b). 
We think that for soliton gas with large density $\rho\gg 1$ the PDF may coincide with the exponential PDF~(\ref{Rayleigh}), however we can neither explain this hypothesis theoretically, nor confirm it numerically. 

Note that, if for different soliton densities $\rho$ we compare PDFs $\mathcal{P}(I)$ of relative wave intensity $I=|\psi|^{2}/\langle\overline{|\psi|^{2}}\rangle$, then at large enough $I$ the PDF is larger for smaller $\rho$, see Fig.~\ref{fig:fig06}(a). 
Thus, the probability of large waves' occurrence is larger for soliton gas with smaller density, if these large waves are counted \textit{relative to the average amplitude of the wavefield} $[\overline{|\psi|^{2}}]^{1/2}$. 
On the other hand, as shown in Fig.~\ref{fig:fig06}(b), the PDF $\mathcal{P}(I_{s})$ of renormalized intensity $I_{s}=|\psi|^{2}/A^{2}$ is almost everywhere larger for larger $\rho$, i.e., the probability of occurrence of waves with the given \textit{absolute intensity} $|\psi|^{2}$ is larger for soliton gas with larger density. 

\begin{figure}[t]\centering
\includegraphics[width=8.5cm]{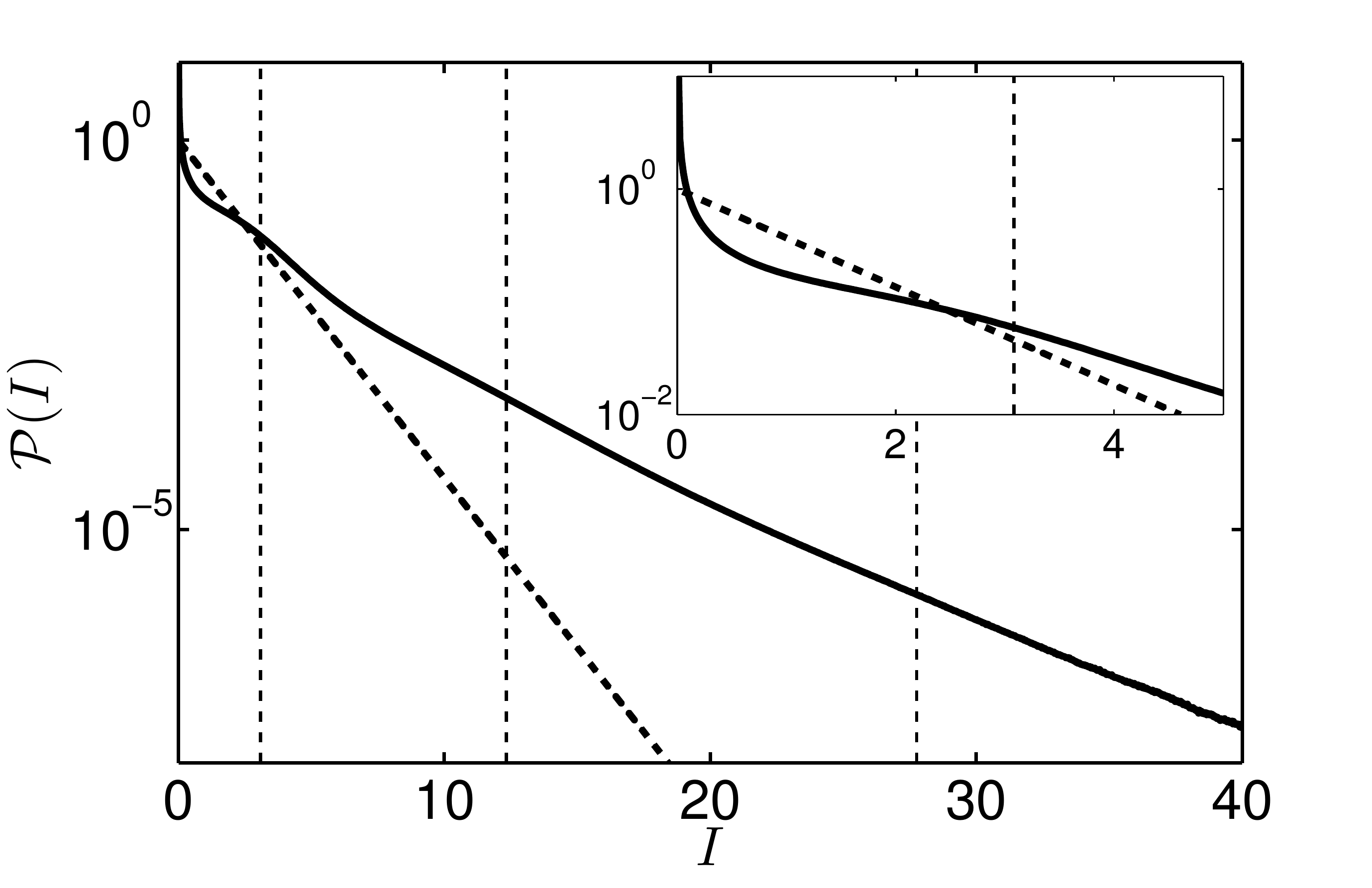}

\caption{\small {\it 
The PDF $\mathcal{P}(I)$ for statistically homogeneous soliton gas with density $\rho=0.32$ and characteristic velocity $V_{0}=2$ containing solitons with Gaussian-distributed amplitudes (thick curve) $a_{n}\sim \mathcal{N}(A,\delta A^{2})$, $\delta A=0.2$, and the exponential PDF $\mathcal{P}_{R}(I)$ (dashed line); we model this soliton gas with $128$-SS. 
The thin vertical lines mark relative intensities $I$ corresponding to amplitudes $A$, $2A$ and $3A$, while the inset shows the PDF at small and moderate $I$. 
}}
\label{fig:fig07}
\end{figure}

The PDF for soliton gas containing solitons of different amplitudes demonstrates the same properties as discussed above, except that the regions of two- and three-soliton collisions are significantly less pronounced, Fig.~\ref{fig:fig07}. 
We model this gas with $128$-SS of density $\rho=0.32$ and with characteristic soliton velocity $V_{0}=2$, using Gaussian distribution for the amplitudes $a_{n}\sim \mathcal{N}(A,\delta A^{2})$, $\delta A=0.2$. 

\begin{figure}[t]\centering
\includegraphics[width=8.5cm]{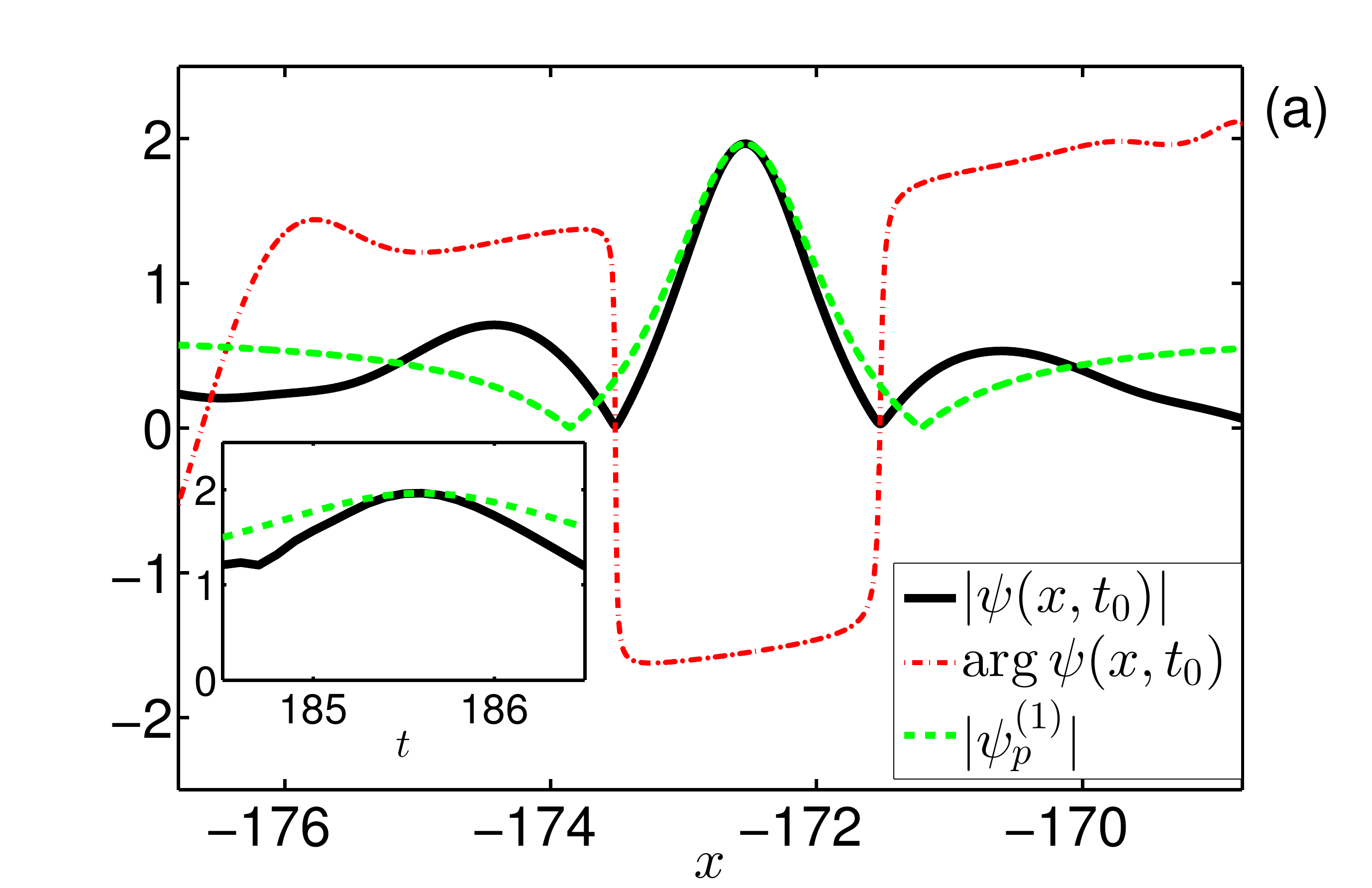}\\
\includegraphics[width=8.5cm]{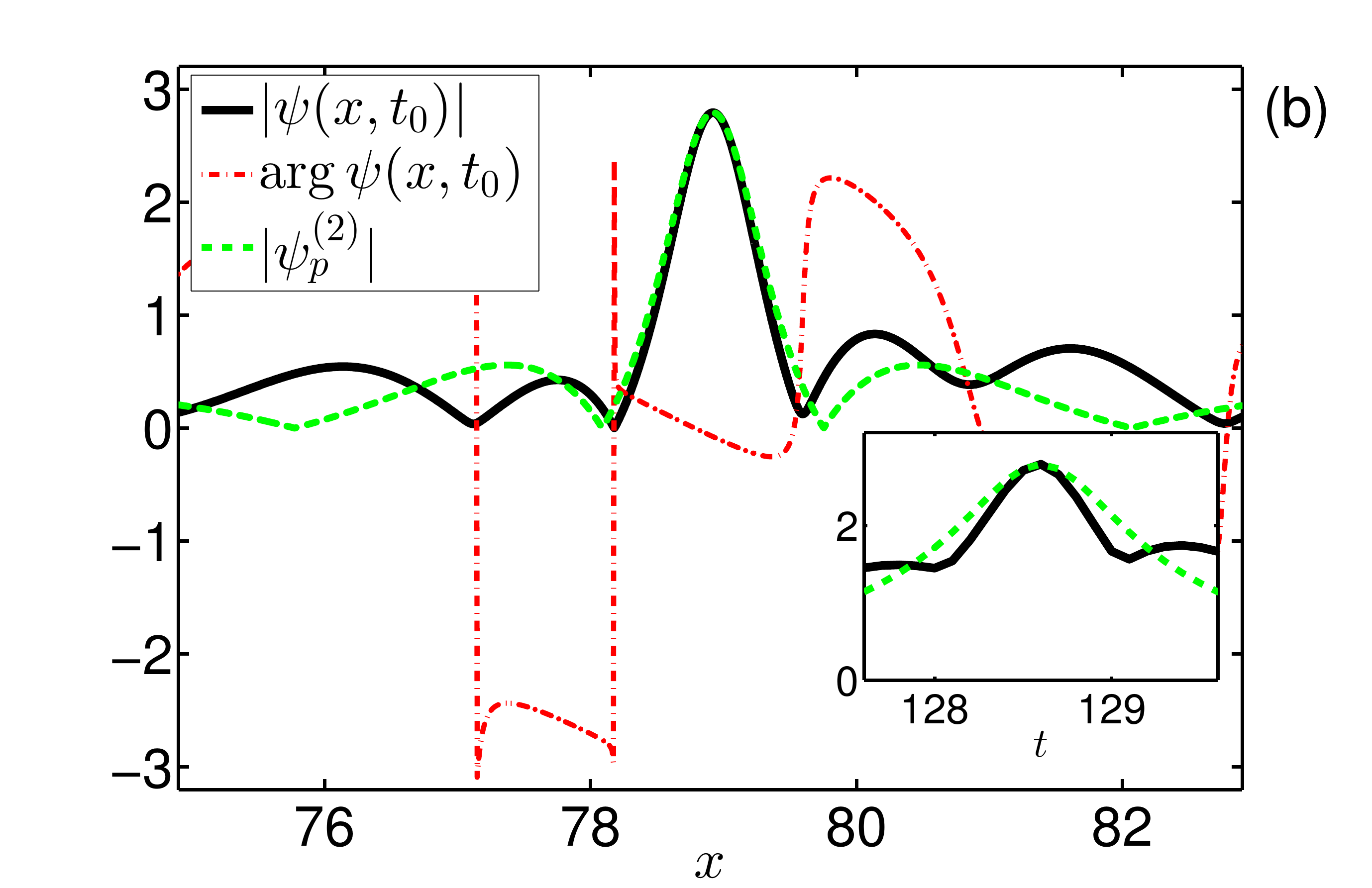}\\
\includegraphics[width=8.5cm]{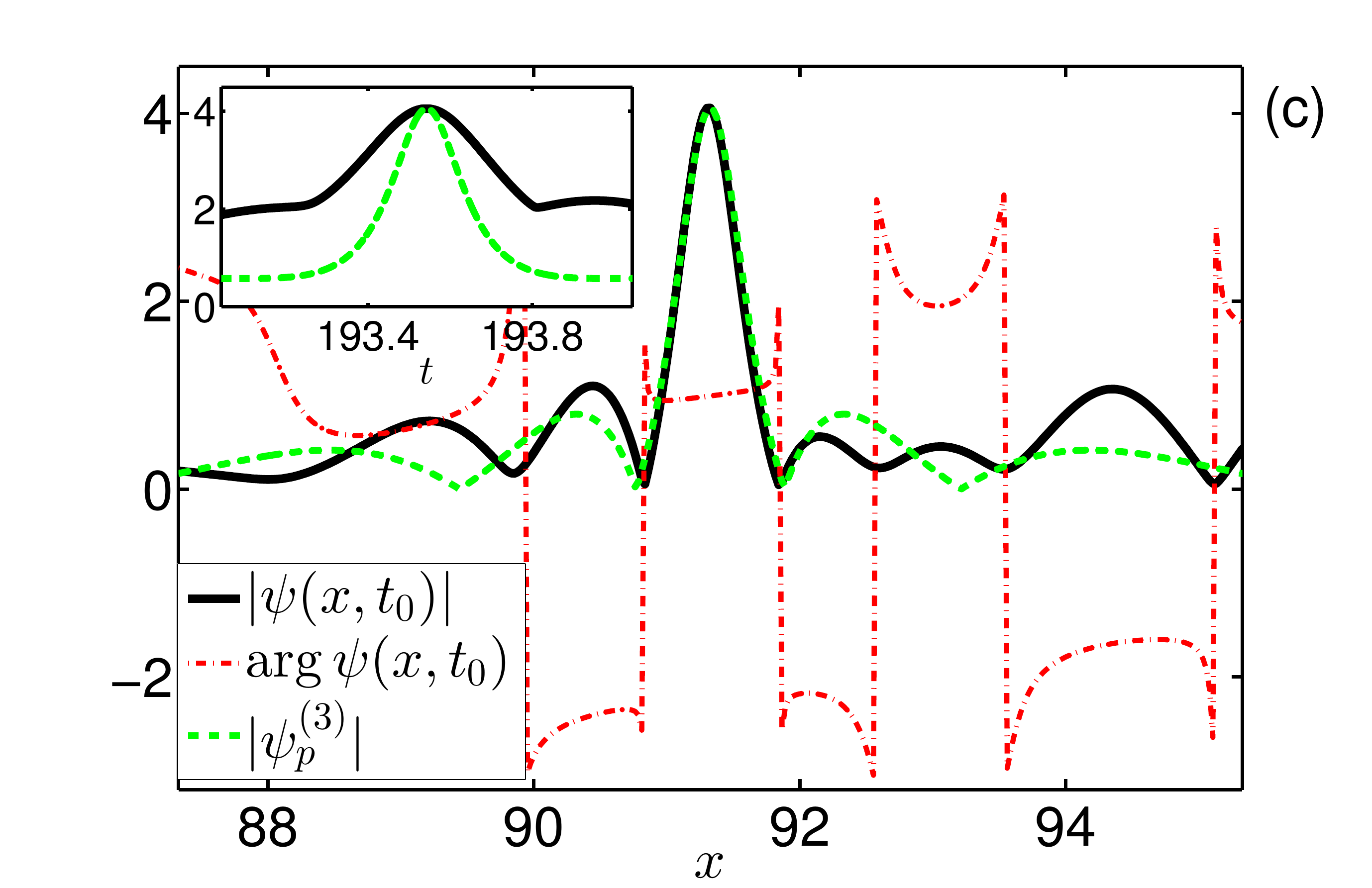}

\caption{\small {\it (Color on-line) 
Large waves at the time $t_{0}$ of their maximal elevation with profiles similar to those of the Peregrine solutions of the first (a), second (b) and third (c) orders. Thick black and dash-dot red show coordinate dependencies for the amplitude $|\psi|$ and phase $\mathrm{arg}\,\psi$, respectively, while dashed green indicates fits by the Peregrine solutions. 
The insets show time-dependency for the maximal amplitude $\max_{x}|\psi|$ (thick black) and its fit with the corresponding Peregrine solution (dashed green). 
}}
\label{fig:fig08}
\end{figure}

\begin{figure}[t]\centering
\includegraphics[width=8.5cm]{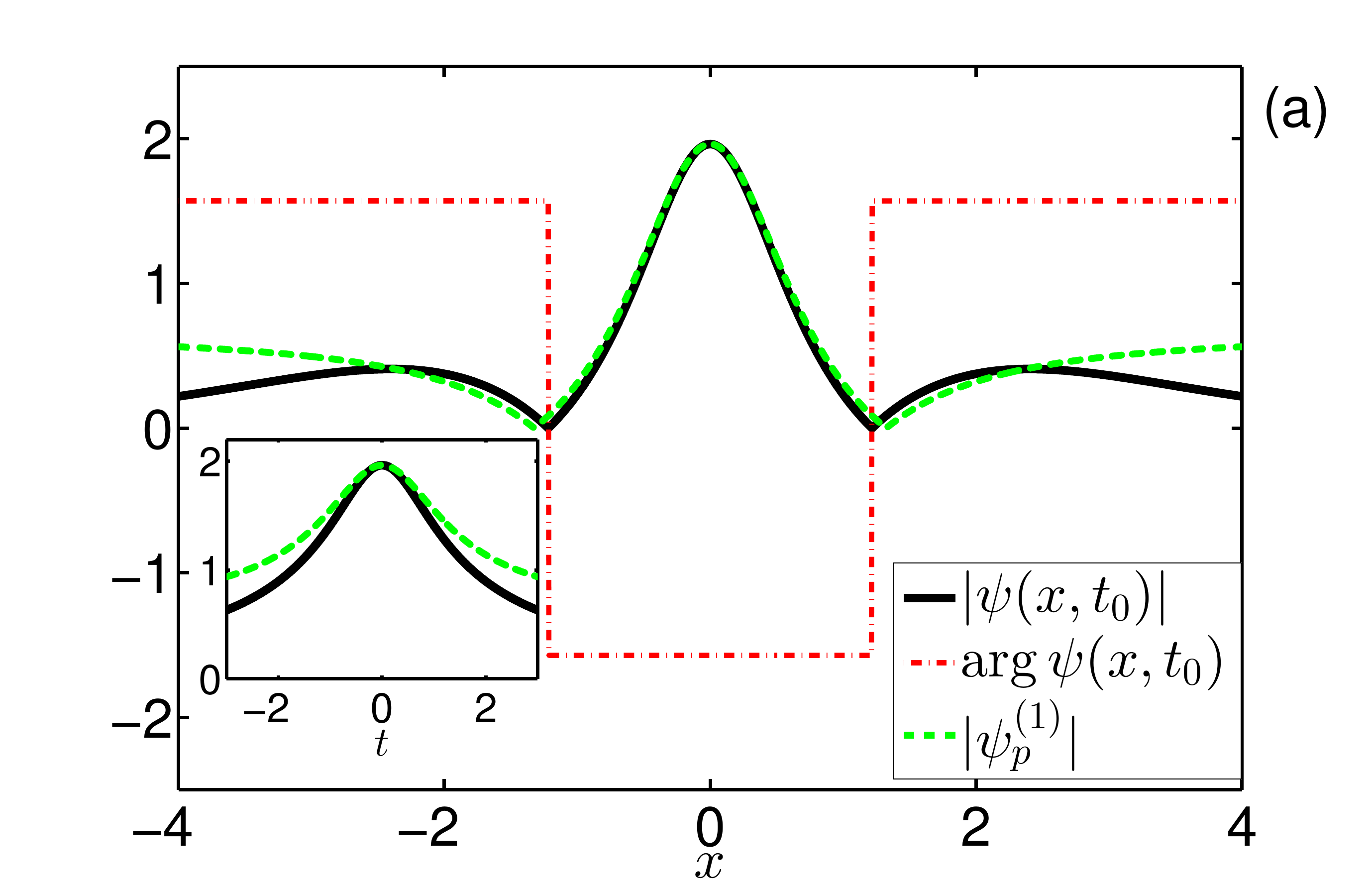}\\
\includegraphics[width=8.5cm]{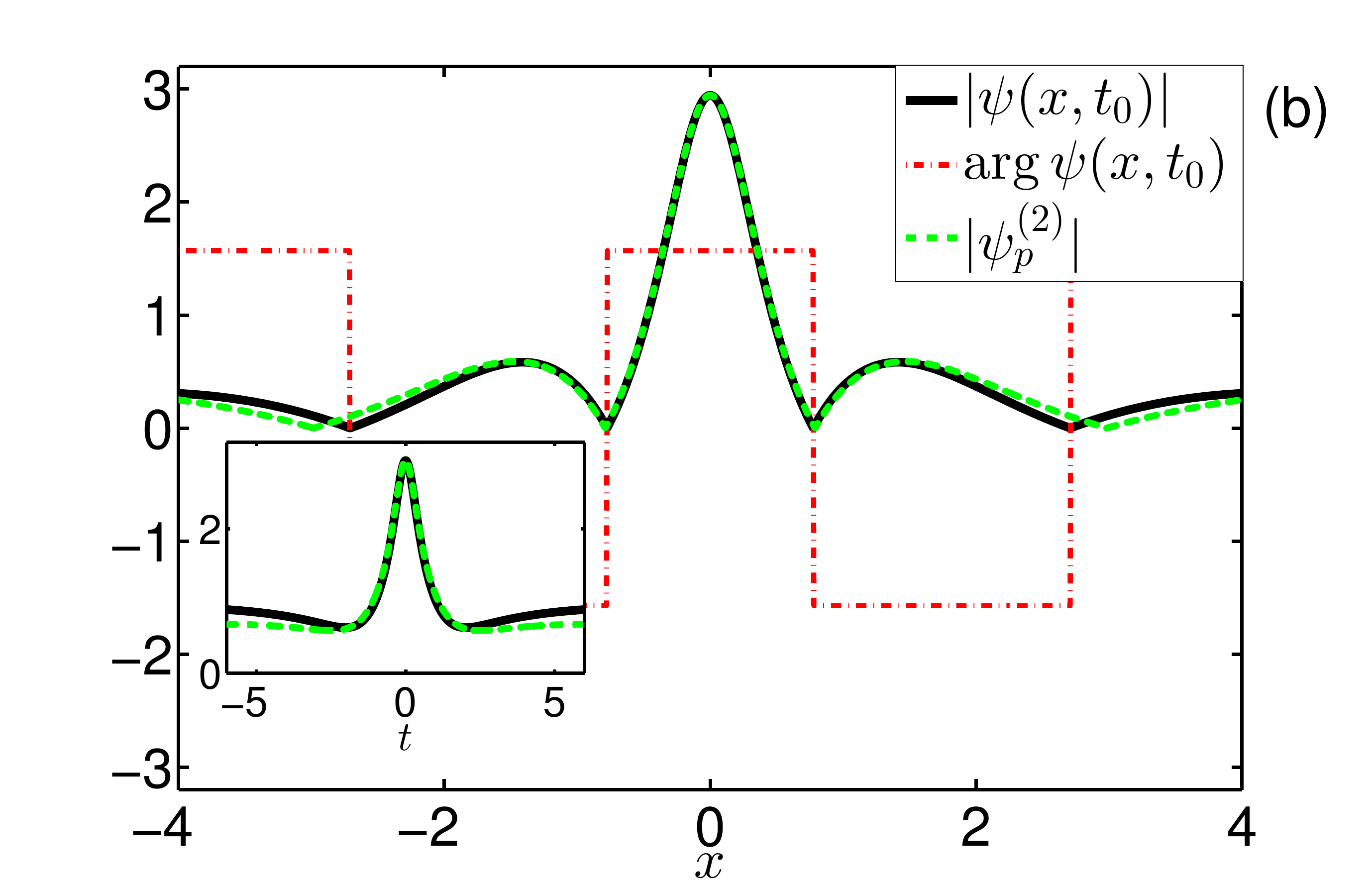}

\caption{\small {\it (Color on-line) 
Specifically designed (a) two-soliton and (b) three-soliton collisions at the time of their maximal elevation $t=0$, and their fits with the Peregrine solution of the first (a) and second (b) orders. 
The colliding solitons have equal amplitudes $a_{n}=A=\pi/3.2$, velocities $v_{1,2}=\pm 0.1$ and symmetric initial positions $x_{01}=-x_{02}$ in figure (a) and $v_{1,3}=\pm 0.1$, $v_{2}=0$,  $x_{01}=-x_{03}$, $x_{02}=0$ in figure (b); the phases $\theta$ of all solitons are equal when they are far away from each other. 
All the notations are the same as in Fig.~\ref{fig:fig08}.
}}
\label{fig:fig09}
\end{figure}

In simulations, we observe rogue waves that exceed soliton amplitude $A$ by up to $4$ times. 
As these waves appear in soliton gas, they are multiple collisions of solitons. 
However, some of these waves, at the time of their maximal elevation, have profiles along $x$-coordinate very similar to those of the so-called Peregrine solutions of the first, second and third orders. 
The Peregrine solution of the first order~\cite{peregrine1983water} is localized in space and time rational solution of the NLS equation~(\ref{NLSE}), 
\begin{equation}\label{PW-p1}
\Psi_{p}^{(1)}(x,t) = e^{i t} \bigg[1 - 4\frac{1+2 i t}{1+4t^{2}+4x^{2}}\bigg]. 
\end{equation}
The Peregrine solutions (also called rational breathers) of the second and third orders are too cumbersome, and we refer the reader to~\cite{akhmediev2009rogue} where they were first found. 
If $\Psi_{p}(x,t)$ is a solution of the NLS equation, then $\psi_{p}=A_{0}\Psi_{p}(X,T)$ with $X = |A_{0}|(x-x_{0})$ and $T = |A_{0}|^{2}(t-t_{0})$ is also a solution; in the case of the Peregrine breathers, $x_{0}$ and $t_{0}$ describe position and time for the maximal elevation of $|\psi_{p}|$. 
Fig.~\ref{fig:fig08} shows examples of large waves which we observe in our simulations, in comparison with the Peregrine solutions of the first, second and third orders, scaled with parameters $A_{0}$, $x_{0}$ and $t_{0}$ to fit the observed waves in their maximal amplitude, and also the position and time of occurrence. 
All large waves in Fig.~\ref{fig:fig08} are obtained using $128$-SS with density $\rho=0.65$, equal amplitudes $a_{n}=A$, and with characteristic soliton velocity $V_{0}=2$. 
The large wave in the figure (a) exceeds the soliton amplitude $A$ by about two times, in figure (b) -- by about three times, and in figure (c) -- by about four times; by checking the temporal dynamics we can confirm that these are instances of two-, three- and four-soliton collisions. 
Along the $x$-coordinate, in the region between the two local minimums closest to the maximal amplitude, the presented large waves are very well approximated by the Peregrine solutions of the first, second and third orders, respectively. 
Note that the phase $\mathrm{arg}\,\psi$ between the two local minimums has almost constant slope with coordinate $x$ for the numerically observed waves, while the Peregrine solutions have constant phase in the same region (i.e., zero slope). 
However, the constant (nonzero) slope can be constructed for the Peregrine solutions as well, if we consider them moving: indeed, if $\Psi_{p}(x,t)$ is a solution of the NLS equation with zero slope for the phase $\mathrm{arg}\,\psi = \mathrm{const}$ in some coordinate region, then $\Psi_{p}(x-vt,t)\,e^{ivx-iv^{2}t/2}$ is also a solution with non-zero slope, $\mathrm{arg}\,\psi - vx = \mathrm{const}$, in the same region. 
Note that the time evolution of the maximal amplitude $\max_{x}|\psi(x,t)|$ for the observed large waves is significantly different from those for the corresponding Peregrine solutions, as shown in the insets in Fig.~\ref{fig:fig08}. 

The large waves presented above demonstrate that quasi-rational profiles similar to those of the Peregrine solutions may appear for rogue waves even they emerge in soliton gas, where all such waves are collisions of solitons. 
This is further demonstrated in Fig.~\ref{fig:fig09}, where we present $2$-SS and $3$-SS with parameters selected to achieve the maximal amplitude growth. 
For the $2$-SS we take velocities $v_{1,2}=\pm 0.1$ and symmetric initial positions $x_{01}=-x_{02}$, while for the $3$-SS -- $v_{1,3}=\pm 0.1$, $v_{2}=0$ and $x_{01}=-x_{03}$, $x_{02}=0$, respectively; solitons have amplitude $A$ and the same value of complex phase $\theta$ when they are far away from each other. 
As shown in the figure, with these parameters the spatial profile for the two-soliton collision at the time of its maximal elevation is very well described by the (scaled) Peregrine solution of the first order, while for the three-soliton collision -- by the Peregrine solution of the second order. 
In the latter case even the temporal behavior of the maximal amplitude $\max_{x}|\psi(x,t)|$ is almost identical to that of the Peregrine solution of the second order. 
In our simulations, colliding solitons do not have ``ideal'' parameters as in Fig.~\ref{fig:fig09}, what may explain larger deviations between the observed large waves in Fig.~\ref{fig:fig08} and their fits with the Peregrine solutions. 
When the soliton parameters are far from the ideal, the spatio-temporal profile of the collision may differ significantly from those of the Peregrine solutions.


\section{Conclusions}
\label{Sec:Conclusions}

In the present paper, for the first time to our knowledge, we have studied statistically homogeneous soliton gas with essential interaction between the solitons. 
As a model, we have used one-dimensional NLS equation of the focusing type; we believe that our methods can be extended straightforwardly to other nonlinear integrable equations. 

At the first step in our study, we created ensembles of $N$-soliton solutions ($N$-SS) with $N\sim 100$, by using the Zakharov-Mikhailov variant of the dressing method applied numerically with $100$-digits precision. 
As far as we are aware, $N$-SS containing so many solitons were not generated by anyone else before. 
Then we put these $N$-SS in a periodic box $L$ and simulated their evolution until the statistically steady state is reached. 
We used this state as a model for homogeneous soliton gas of density $\rho\propto N/L$ in an infinite space; we confirmed that for sufficiently large $N$ and $L$ our results depend on them only in combination $N/L$. 
We examined the major statistical characteristics of soliton gas, in particular the kinetic and potential energies, the kurtosis, the wave-action spectrum and the PDF of relative wave intensity, depending on soliton density and velocities. 

We have shown that in the case of rarefied soliton gas $\rho\ll 1$ the kinetic and potential energies, as well as the kurtosis, are very well described by analytical relations~(\ref{raregas}) derived without taking into account soliton interactions. 
For larger density $\rho$ and characteristic soliton velocity $V_{0}$, we observe increasing next-order corrections leading to increased absolute values for all these three characteristics. 
These next-order corrections come from enhanced soliton interactions due to decreased spacing and more frequent collisions, respectively. 
The wave-action spectrum for soliton gas is smooth, decays close to exponentially at large wavenumbers and widens with increasing $\rho$ and $V_{0}$. 

The PDF of relative wave intensity has the form of a composition of PDFs representing a singular soliton and soliton interactions. 
Compared to the cnoidal wave initial conditions, the PDF deviates from the exponential (Rayleigh) distribution~(\ref{Rayleigh}) much more pronouncedly, especially at the region of soliton interactions where it exceeds the exponential PDF by orders of magnitude. 
This excess is larger for soliton gas with larger velocities, what corresponds to more frequent soliton collisions. 
For rarefied soliton gas $\rho\ll 1$, the average amplitude of the wavefield is much smaller than the soliton amplitude and the PDF deviates from the exponential PDF drastically. 
For larger densities, solitons interact stronger and the PDF transforms closer to the exponential distribution. 
We think that for dense soliton gas $\rho\gg 1$ the PDF may match the exponential one, what is supported by the behavior of the kurtosis approaching to $2$ with increasing density. 
Soliton gas containing solitons of different amplitudes demonstrate the similar properties, except that the regions of soliton interactions on the PDF are less pronounced. 

Rogue waves emerging in soliton gas are collisions of solitons, and some of these collisions have spatial profiles very similar to those of the (scaled) Peregrine solutions of different orders. 
In particular, we present specifically designed examples of two- and three-soliton collisions, which have almost the same spatial profiles as the Peregrine solutions of the first and second orders. 
In the case of the three-soliton collision, even the temporal dependency of the maximal amplitude is very well approximated by that of the Peregrine solution of the second order. 
When soliton parameters are far from the ``ideal'' sets, the emerging large waves differ significantly from the rational breathers. 
In our opinion, these results highlight that similarity in spatial and/or temporal behavior cannot be used to draw conclusions on rogue waves' composition and origin. 

For a statistical study, it is crucial to define the ensemble of initial conditions. 
In this paper, we have used initial conditions with fixed value of wave action (average intensity) and with zero momentum, while the integrals of higher order were not fixed; for instance, the total energy could change significantly from one realization to another. 
To check the influence of this effect, we examined soliton gas for which -- in addition to the wave action and the momentum -- the value of the total energy was also fixed, and came to the identical results. 

We suggest that our methods for generation of initial conditions from known scattering data can be used to examine turbulence governed by other integrable equations and developing from other types of initial conditions, e.g. containing nonlinear dispersive waves and different types of breathers~\cite{frumin2017new,gelash2018formation}. 
We believe that, in general, our approach can be promising, as it allows to study turbulence with controlled initial conditions, i.e., with exact knowledge which nonlinear objects interact during the evolution. 
Our methods can also be used in optical fibre communications, where strongly interacting $N$-SS were recently proposed as information carrier~\cite{frumin2017new}. 

\begin{center}
\textbf{Acknowledgements.}
\end{center}

Simulations were performed at the Novosibirsk Supercomputer Center (NSU). 
The work described in Section~\ref{Sec:NumMethods}(a) was supported by the Russian Science Foundation Grant No. 17-71-10128 to A.G., the other parts of the study were supported by the Russian Science Foundation Grant No. 14-22-00174.

\end{document}